# A Bayesian modelling framework to quantify multiple sources of spatial variation for disease mapping

Sophie A. Lee[1,2*], Theodoros Economou[3], Rachel Lowe[1,2,4,5]

1. Centre for Mathematical Modelling of Infectious Diseases, London School of Hygiene & Tropical Medicine, London, UK
2. Centre on Climate Change and Planetary Health, London School of Hygiene & Tropical Medicine, London, UK
3. Climate and Atmosphere Research Centre, The Cyprus Institute, Nicosia, Cyprus
4. Barcelona Supercomputing Center (BSC), Barcelona, Spain
5. Catalan Institution for Research and Advanced Studies (ICREA), Barcelona, Spain

*Corresponding author
Email: sophie.lee@lshtm.ac.uk

# Abstract

Spatial connectivity is an important consideration when modelling infectious disease data across a geographical region. Connectivity can arise for many reasons, including shared characteristics between regions, and human or vector movement. Bayesian hierarchical models include structured random effects to account for spatial connectivity. However, conventional approaches require the spatial structure to be fully defined prior to model fitting. By applying penalised smoothing splines to coordinates, we create 2-dimensional smooth surfaces describing the spatial structure of the data whilst making minimal assumptions about the structure. The result is a non-stationary surface which is setting specific. These surfaces can be incorporated into a hierarchical modelling framework and interpreted similarly to traditional random effects. Through simulation studies we show that the splines can be applied to any continuous connectivity measure, including measures of human movement, and that the models can be extended to explore multiple sources of spatial structure in the data. Using Bayesian inference and simulation, the relative contribution of each spatial structure can be computed and used to generate hypotheses about the drivers of disease. These models were found to perform at least as well as existing modelling frameworks, whilst allowing for future extensions and multiple sources of spatial connectivity.

# 1. Introduction

When modelling infectious disease data across a geographical region, it is important to account for potential spatial connectivity between areas. For example, spatial connectivity may arise from human or vector movement contributing to the spread of a vector-borne disease, or unobservable climatic, behavioural, biological, and socioeconomic factors shared between areas. Conventionally, Bayesian hierarchical models aim to account for this spatial connectivity by including spatially structured random components within the model [1–3]. Fully Bayesian modelling approaches require the spatial structure of components to be defined prior to model-fitting. However, the spatial structure of the data may not be fully known. A recent systematic review found that all Bayesian hierarchical models for mosquito-borne diseases used a distance-based spatial structure, assuming connectivity between regions only exists between neighbours or close observations [4].

Spatial autocorrelation in disease count data may be attributable to multiple sources of connectivity. For example, dengue incidence is associated with climate variation, vector control interventions and levels of immunity in the population which is likely shared between close regions [5]. However, dengue is also influenced by human movement which creates links between distant regions that a distance-based spatial connectivity assumption would not capture [6,7]. In this example, multiple random terms would be required within a Bayesian hierarchical model to capture the different sources of connectivity and quantify the relative importance of each to the disease transmission process.

In this paper, we present a Bayesian hierarchical modelling framework that uses penalised smoothing splines as a flexible method for structuring spatial model components. Smoothing splines use data to inform spatial components, given smoothing assumptions, rather than requiring the full specification of the spatial structure prior to model fitting [8,9]. The result is a non-stationary structure which is setting-specific and requires minimal user assumptions. This approach allows multiple spatially structured random components to be incorporated into the same model and can distinguish between these structures to quantify their relative contribution to the overall spatial structure. Although this study focuses on disease mapping models of count data, we also show that this method can be used for models of binary data.

## 2. Modelling approach

### 2.1 Disease mapping

Disease mapping is an important statistical tool used in epidemiology to explore spatial variation in disease incidence rates. Disease mapping models can generate and test hypotheses about associations between disease and a variety of potential explanatory variables, such as environmental and socio-economic factors [2,10]. Typically, disease counts, $y_i: (i = 1, \ldots n)$, are collected across a study area separated into *n* contiguous areas. These counts are combined with an offset $\log(\xi_i)$ describing the underlying population at risk is each area $i$. For instance, $y_i/\xi_i$ is the empirical incidence rate in $i$ when $\xi_i$ is population count. Where a disease is rare or areas within the study are small, estimates of the incidence are highly uncertain and thus unstable and inflated. To overcome this issue, Bayesian (hierarchical) modelling approaches have been developed to allow information from connected regions to be included in the rate estimation using random effects (data pooling). Conventionally, these models take the form:

$$y_i \sim p(E(y_i), \psi)$$

$$log(E(y_i)) = log(\xi_i) + \alpha + S_i \quad (1)$$

where $p$ is a suitable count distribution (e.g. Poisson, Negative Binomial), $E(y_i)$ is the expected count, $\alpha$ is the intercept or baseline risk, $S_i$ are spatially structured random components and $\psi$ are hyperparameters of the distribution. The definition of $S_i$ (which describes the spatial structure of $E(y_i)$ on the log scale, after correcting for $\xi_i$) depends on the disease of interest and the assumed spatial structure in the data. A recent systematic review found that spatial statistical models used to study mosquito-borne diseases only considered distance-based connectivity when defining the structure of such spatial random effects [4]. The most common spatial structure assumed connectivity between regions if and only if they share a border using a conditional autoregressive (CAR) model:

$$S_i | S_{j \neq i} \sim N\left(\frac{\sum_{j \neq i} W_{ij} S_j}{\sum_{j \neq i} W_{ij}}, \frac{\sigma_s^2}{\sum_{j \neq i} W_{ij}}\right) \quad (2)$$

where $W_{ij}$ are proximity weights, often defined as $W_{ij} = 1$ if *i* and *j* share a border, and 0 otherwise. Although the conditional independence assumption intrinsic to neighbourhood-based spatial structures allows for efficient Bayesian computation [11], the nature of spatial connectivity is likely to be more complex and differ across settings. A smooth function with a structure defined using the data rather than prior to model fitting provides a flexible alternative and allows spatial dependency structures to be specific to each setting.

2.2 Penalised smoothing splines

Smoothing splines, or smooth functions, are used in generalised additive models (GAMs) to explore nonlinear relationships between a response variable and one or more covariate(s). Smoothing splines are constructed as a linear combination of basis functions, $b_j$ (functions applied to the covariate(s) at given intervals, determined by the type of smoothing spline chosen), multiplied by regression coefficients, $\beta_j$ [8]. For example:

$$f(x) = \sum_{j=1}^{K} \beta_j b_j(x) \quad (3)$$

Where $f$ is a smooth function (the smoothing spline), $x$ is the covariate of interest, and *K* is the number of 'knots', or turning points, in the smooth function. The number of knots should be chosen to be large enough that the smooth function adequately describes the data, but not so large that they overfit or become 'overly wiggly'. To achieve this, a smoothing penalty parameter, $\lambda$, is introduced and estimated using the data to avoid over-fitting when *K* is too large (e.g., as $\lambda \to \infty$, $f(x)$ becomes linear) [9].

Regression coefficients $\boldsymbol{\beta}$ are estimated using restricted maximum likelihood (REML) which imposes a smoothing penalty on the coefficients of the form:

$$\lambda \boldsymbol{\beta}^T P \boldsymbol{\beta} \quad (4)$$

where $\lambda$ is the penalty parameter introduced earlier, and $P$ is a penalty matrix computed prior to model fitting (based on the type of smoothing spline chosen) [8,9]. The penalty parameter, matrix, and basis functions can be estimated efficiently using the mgcv package [12]. Although the mgcv package uses empirical methods to estimate the parameters defining smoothing splines, the results can be interpreted from a Bayesian perspective.

## 2.3 Bayesian interpretation of penalised smoothing splines

The assumption that smoothing functions $f$ are more smooth than wiggly can be considered a prior belief on the values that the coefficients can take. This prior can be formalised and incorporated into Bayesian inference by assuming the regression coefficients $\beta$ have the prior distribution:

$$\beta \sim N(0, P^-/\lambda) \quad (5)$$

where $P^-/\lambda$ is the covariance matrix [8,9]. However, the precision matrix $P\lambda$ is rank-deficient so is instead replaced by $P_0\lambda_0 + P_1\lambda_1$ where the first term relates to a penalty on the null space of the smooth function and the second is the wiggliness penalty [13]. The interpretation of this is that the penalty matrix is separated into penalised components through $P_1$ (relating to wiggly behaviour) and non-penalised components through $P_0$. The splines $b_j(x)$ and penalty matrices can be efficiently generated using the jagam function in the mgcv package [13]. The definition of smoothing splines as linear combinations of (known) basis functions and (unknown) coefficients means that they can be entered into hierarchical models [14] and implemented using Bayesian inferential methods such as Markov chain Monte Carlo (MCMC). Under these conditions, the resulting penalised smoothing splines can be interpreted as random effects [8,15].

## 2.4 Spatial smoothing splines within Bayesian hierarchical models

In this study, we applied penalised smoothing splines to coordinates describing the relative 'connectivity' of regions (e.g. coordinates of the centroid of regions). This created 2-

dimensional smooth surfaces describing spatial patterns in the data. Thin plate regression splines are relatively efficient at estimating smooths over multiple variables and do not require a surface to be stationary. In addition, thin plate regression splines have low posterior correlation between parameters which improves mixing when using MCMC methods [16,17]. If a coordinate system does not currently exist that describes the connectivity in question, this can be created from a continuous measure using multidimensional scaling (MDS). MDS translates a continuous measure of 'distance' or connectivity between observations onto an abstract cartesian space and returns a set of coordinates [18]. For example, when connectivity is assumed to arise due to human movement, this could be defined as a continuous measure such as the number of air travel passengers, or an estimate from a movement model, such as a gravity or a radiation model [19,20], which assumes the number of people moving between areas is a function of population and distance.

Smooth surfaces were defined using splines and included in Bayesian hierarchical models of count data using the procedures detailed above. Models were implemented using NIMBLE [21,22], a flexible programme that implements Bayesian models created in the BUGS language using MCMC methods within R [23]. The flexibility of this framework means that multiple spatial smooth surfaces can be included in the same model with different connectivity assumptions (e.g. distance-based and human movement). Interpreting the smooth surfaces over the various connectivity measures as random, means the relative contribution of each spatial structure can be quantified by calculating the proportion of the overall variance of the random terms that is captured by each spatial term.

## 3. Simulation study 1: a single source of spatial structure

In this section, we present a simulation study in which we apply Bayesian spatial models to data generated from a distance-based spatial structure. We compare model performance between the penalised regression spline approach and a neighbourhood-based CAR model. A further simulation study assuming a single source of human movement-based connectivity is presented in the supplementary materials.

## 3.1 Data generation:

Fictitious disease count data was generated from a Poisson distribution for each of the 1,013 municipalities in South Brazil, the region used in the case study (Section 5), from model (1). The log of the population divided by 100,000, $log(\xi_i)$, was included as an offset (Figure S1). The population of each municipality was taken from the Brazilian census and described in Section 5.1. The intercept term $\alpha$ was set to zero, while the term $S_i$ was defined by

$$S_i = \sqrt{\phi} \cdot sm(x_i, z_i) + \sqrt{(1-\phi)} \cdot \varepsilon_i \quad (7)$$

where $\phi$ is a mixing parameter, taking values between 0 and 1, which measures the contribution of each term (if we interpret $sm(x_i, z_i)$ as random and independent of $\varepsilon_i$) to the overall variance of $S_i$, and $\varepsilon_i \sim N(0,1)$. $sm(x_i, z_i)$ is a continuous function applied to connectivity coordinates $(x_i, z_i)$ to emulate a spatially structured surface (Figure 1b, taken from [24]):

$$sm(x,z) = \pi \sigma_x \sigma_z \left( 1.2 e^{-(x-0.2)^2/\sigma_x^2 - (z-0.3)^2/\sigma_z^2} + 0.8 e^{\frac{-(x-0.7)^2}{\sigma_x^2} - \frac{)z-0.8)^2}{\sigma_z^2}} \right) \quad (8)$$

$$\sigma_x = 0.3, \quad \sigma_z = 0.4$$

To create a distance-based spatial structure, the smooth function $sm$ was applied to coordinates of the centroid of municipalities which were scaled to take values between 0 and 1. The function $sm(x_i, z_i)$ was centred at 0 by subtracting the overall mean from each value. 11 simulated data sets were produced using equation (7), setting values of $\phi$ between 0 and 1 at intervals of 0.1 (Figure 1).

a)

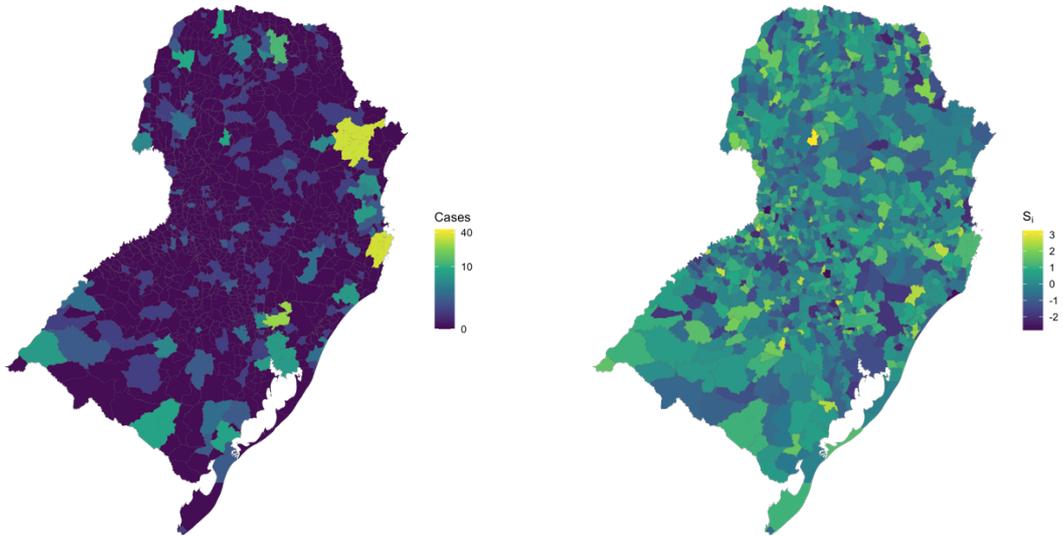

b)

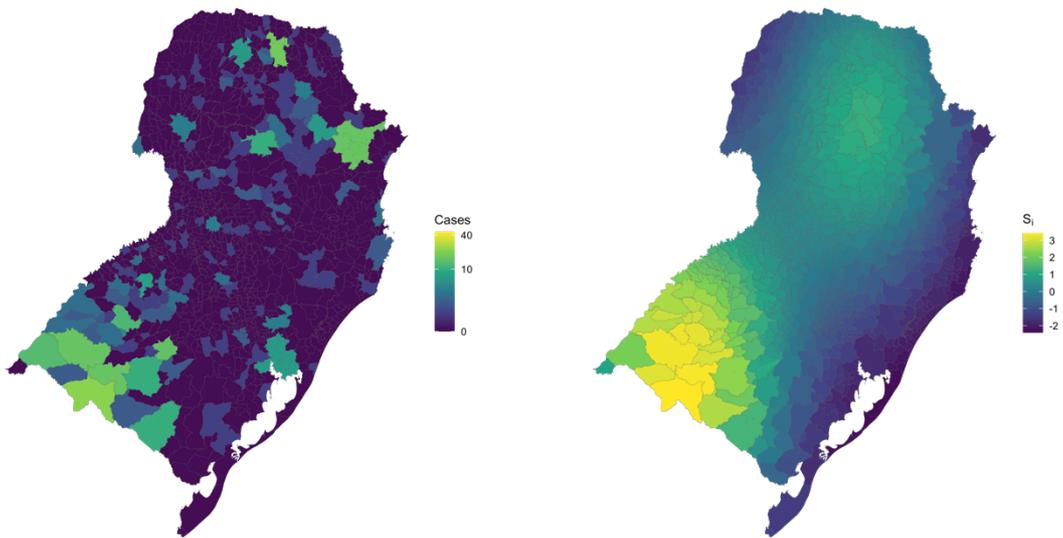

c)

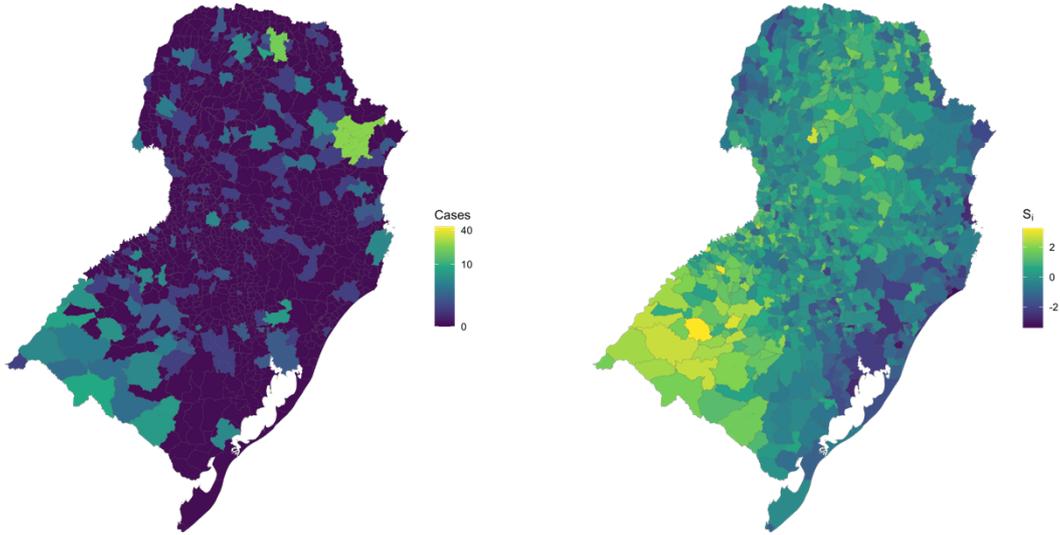

**Figure 1: Simulated disease counts (left) and spatial random effects (right) under a distance-based structure using different spatial structure combinations.** The number of cases simulated from a Poisson model and the underlying spatial structure where the data has a) no spatial structure ($\phi = 0$), b) a distance-based structure only ($\phi = 1$), and c) equal contribution of both structures ($\phi = 0.5$).

## 3.2 Modelling approach

Two Poisson models containing spatially structured and unstructured random components were applied to each simulated dataset:

$$y_i \sim Poisson\big(E(y_i)\big)$$

$$log\big(E(y_i)\big) = log(\xi_i) + \alpha + u_i + v_i \quad (9)$$

$$log\big(E(y_i)\big) = log(\xi_i) + \alpha + \frac{1}{\tau}\big(\sqrt{\phi}u_{*i} + \sqrt{1-\phi}v_{*i}\big) \quad (10)$$

In model (9), $u_i$ is a spatially structured term, constructed using a thin plate regression spline on the coordinates of the centroid of each municipality, and $v_i$ is a spatially unstructured term, assumed to follow a zero-mean normal distribution, representing heterogeneity between regions. This spatial smooth model was compared to a more conventional random effect approach based on R-INLA's BYM2 model, which is often used to capture spatial structure in disease mapping [3,25,26]. In model (10), $u_{*i}$ are spatially structured random effects assuming a CAR model with a binary neighbourhood matrix (see equation (2)), $v_{*i}$ are unstructured Normal random effects, and $\phi$ is a mixing parameter, measuring the contribution of each random effect to the marginal variance $\left(\frac{1}{\tau^2}\right)$ of the overall random effect [3,25]. Here, $\phi = 1$ represents a purely spatial model, equivalent to an intrinsic CAR model [27], and $\phi = 0$ indicates no spatial structure in the data. Spatial smooth models were fitted using MCMC simulations in R via the NIMBLE package [21], and random effect models were created using integrated nested Laplace approximations (INLA) [3,28].

Model comparison was based on mean absolute error (MAE) and WAIC, an information criterion used to assess the predictive accuracy of Bayesian models [29]. Lower values of MAE and WAIC are preferred. The relative contribution of the spatially structured term, $u_i$, to the overall random terms in the spatial smooth model was defined as the proportion of the overall random term variance explained by $u$ ($var(u)/var(u + v)$). This was estimated using samples from the posterior distribution of $u$ and $v$. We compared estimates of the $\phi$ hyperparameter from INLA, the relative contribution of $u_i$ to the random effect variance from NIMBLE, and the known proportion of spatial variance used in the simulation. All analyses were carried out using R version 4.1.1 [23]. Code used to simulate data and perform analyses is available here: https://github.com/sophie-a-lee/spatial_smooth_framework.

### 3.3 Results

Mean absolute errors and WAIC values show that model performance was similar between the smoothing spline and BYM2 models (Table 1). The WAIC showed the smoothing spline model performed slightly better on all simulated datasets apart from one, although the MAE preferred the BYM2 models. This illustrates that the smoothing spline was able to detect spatial connectivity between neighbouring regions whilst being flexible enough to capture alternative structures. The 95% credible interval of the intercept coefficient estimate contained the true value 0 for all models besides one for both approaches (Figure S2).

**Table 1: Model comparison statistics and mean estimates of the mixing parameter, $\phi$, from the smoothing spline and INLA BYM2 models**. Mean absolute error (MAE) and WAIC calculated for the spatial spline and BYM2 models for each simulated dataset. The lowest MAE and WAIC, and the $\phi$ estimate closest to the value used in each simulation are highlighted in bold.

| | Smoothing spline model | | | INLA BYM2 model | | |
|---|---|---|---|---|---|---|
| $\phi$ | MAE | WAIC | $\phi$ estimate | MAE | WAIC | $\phi$ estimate |
| 0 | 1.51 | **996.94** | **0.041** | **1.04** | 1005.79 | 0.072 |
| 0.1 | 1.54 | **1030.64** | **0.121** | **1.11** | 1034.29 | 0.279 |
| 0.2 | 1.33 | 932.42 | **0.26** | 1.04 | **931.79** | 0.486 |
| 0.3 | 1.27 | **909.42** | **0.253** | 0.93 | 912.5 | 0.572 |
| 0.4 | 1.39 | **961.67** | **0.375** | 1.08 | 976.12 | 0.625 |
| 0.5 | 1.54 | **935.09** | **0.601** | 1.21 | 954.34 | 0.668 |
| 0.6 | 1.5 | **881.09** | **0.512** | 1.13 | 973.61 | 0.757 |
| 0.7 | 1.45 | **931.85** | **0.641** | 1.17 | 989.24 | 0.89 |
| 0.8 | 1.63 | **947.51** | **0.808** | 1.37 | 983.96 | 0.951 |
| 0.9 | 1.59 | **876.37** | **0.918** | 1.37 | 922.29 | 0.963 |
| 1 | 1.48 | **875.42** | 0.797 | **1.25** | 924.14 | **0.948** |

We found that the spatial spline model estimates were closer to the true value of $\phi$ than the BYM2 model for most simulations (Figure 2, Table 1), and that INLA's estimates of this parameter was not always consistent with the true value. This indicates that the spatial spline models were able to identify and quantify the relative contribution of this spatial structure within the data as well as (if not better than) INLA's BYM2 models.

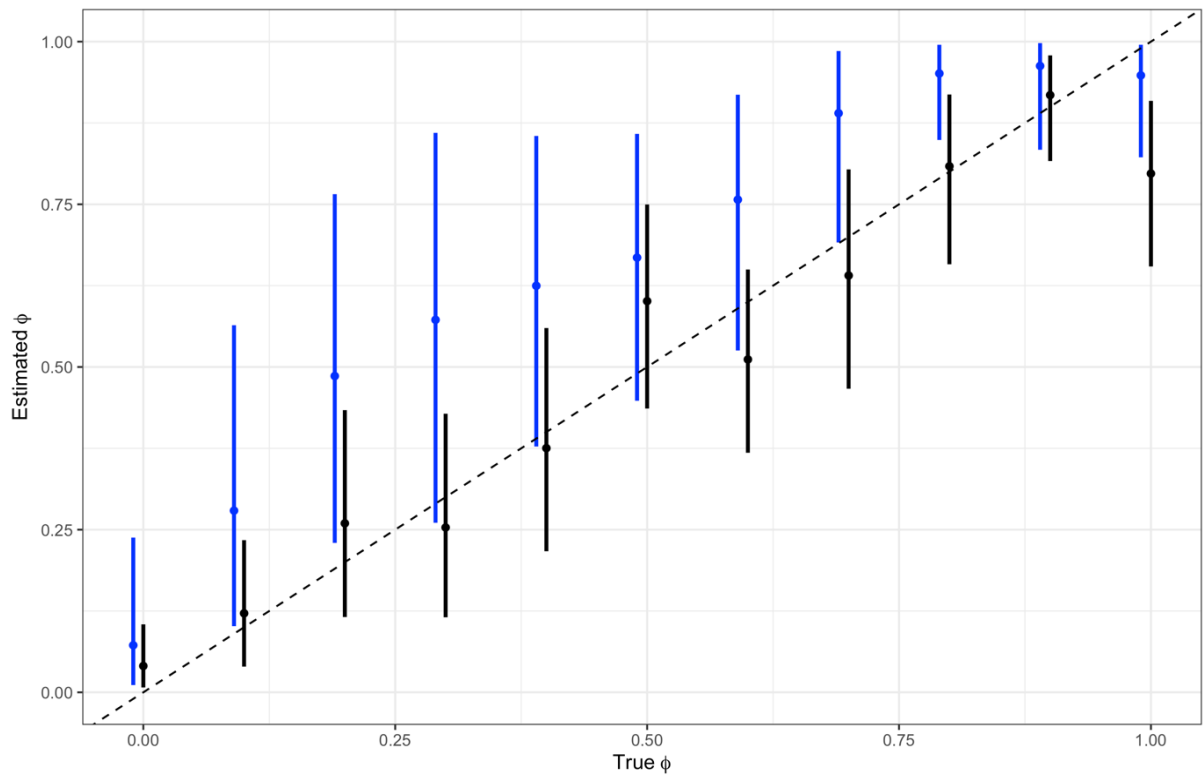

Figure 2: The mean and 95% credible interval of estimated $\phi$ values extracted from models including a smoothing spline (black) and BYM2 (blue) compared to the known value (dashed line). Estimated $\phi$ values for the smoothing spline model were calculated using the proportion of the random effect variance explained by the spatially structured term and were extracted from INLA output for the BYM2 model.

## 4. Simulation study 2: two sources of spatial structure

In this section, we present another simulation study in which we apply Bayesian spatial models to data generated with two sources of spatial connectivity: distance-based and human movement-based.

### 4.1 Data generation

An extension of the spatial term, $S_i$, in equation (7) was used to generate data with spatial connectivity arising from two different sources:

$$S_i = \sqrt{\phi_1} \cdot sm(a_i, b_i) + \sqrt{\phi_2} \cdot sm(c_i, d_i) + \sqrt{\phi_3} \cdot \varepsilon_i \quad (11)$$

$$\phi_1 + \phi_2 + \phi_3 = 1$$

Where $sm$ is a smooth function (equation (8)), applied to coordinates describing distance-based connectivity $(a_i, b_i)$, and human movement-based connectivity $(c_i, d_i)$. The coordinates of the centroid of municipalities were scaled to take values between 0 and 1 and used to describe distance-based connectivity $(a_i, b_i)$. As a coordinate system describing connectivity arising from human movement does not exist, we applied MDS to an estimate of the number of people moving between municipalities, generated using a movement model described in the supplementary materials, to create coordinates $c_i$ and $d_i$ (Figure S3).

In this example, we used three scaling parameters, $\phi_1$, $\phi_2$ and $\phi_3$, to describe the relative contribution of each random term to the marginal variance. We held $\phi_3$ constant at 0.1, with $\phi_1$ and $\phi_2$ taking values between 0 and 0.9 at intervals of 0.1, creating 10 simulated datasets (Figure 3).

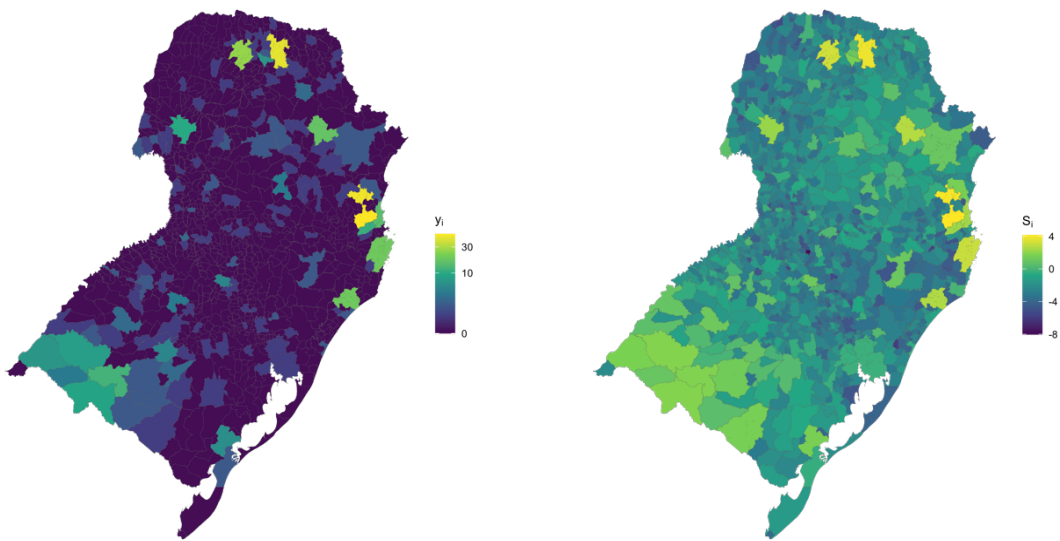

**Figure 3: Simulated data containing two sources of spatial structure.** Simulated disease counts, $y_i$ (left), and spatial random terms, $S_i$ (right), for South Brazil generated using equation (4), where $\phi_1 = 0.4$, $\phi_2 = 0.5$, and $\phi_3 = 0.1$.

## 4.2 Modelling approach

We applied a Poisson spatial model to each simulated dataset which contained three random terms:

$$y_i \sim Poisson(E(y_i))$$

$$log(E(y_i)) = log(\xi_i) + \alpha + u_{1,i} + u_{2,i} + v_i \quad (12)$$

Where $u_{1,i}$ is constructed using a thin plate regression spline applied to coordinates of the centroids of municipalities, and $u_{2,i}$ is structured using a thin plate regression spline applied to human movement-based connectivity coordinates described previously. $v_i$ is assumed to have no spatial structure and represents unobserved heterogeneity between municipalities. We compared the proportion of the marginal variance explained by each random term and compared these to the known $\phi$ values used in data generation.

## Results

We found that the models were able to accurately estimate the intercept coefficient value of 0 across most simulated datasets (Figure S4). Estimates of the relative contribution of each random term to the overall spatial structure were close to $\phi$ values used in simulations and were able to detect the increasing contributions of distance-based and human movement-based terms as the true value increased (Figure 4).

a)

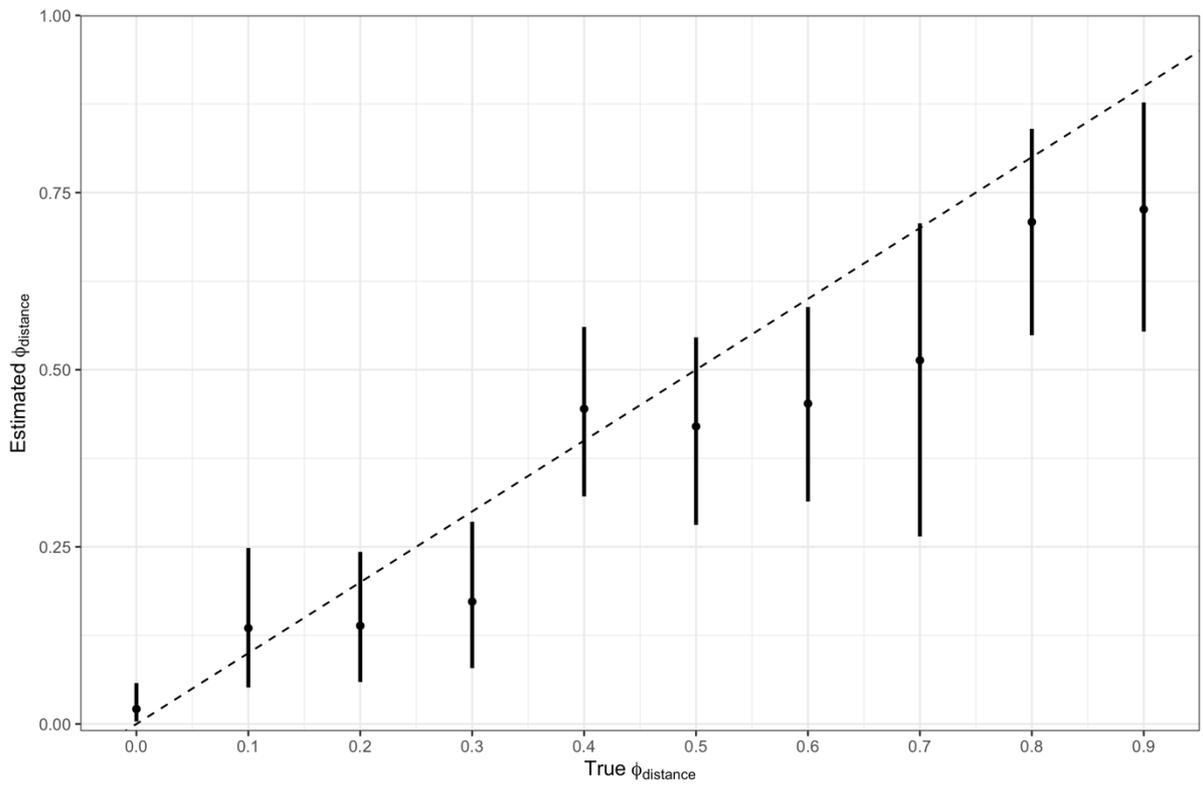

b)

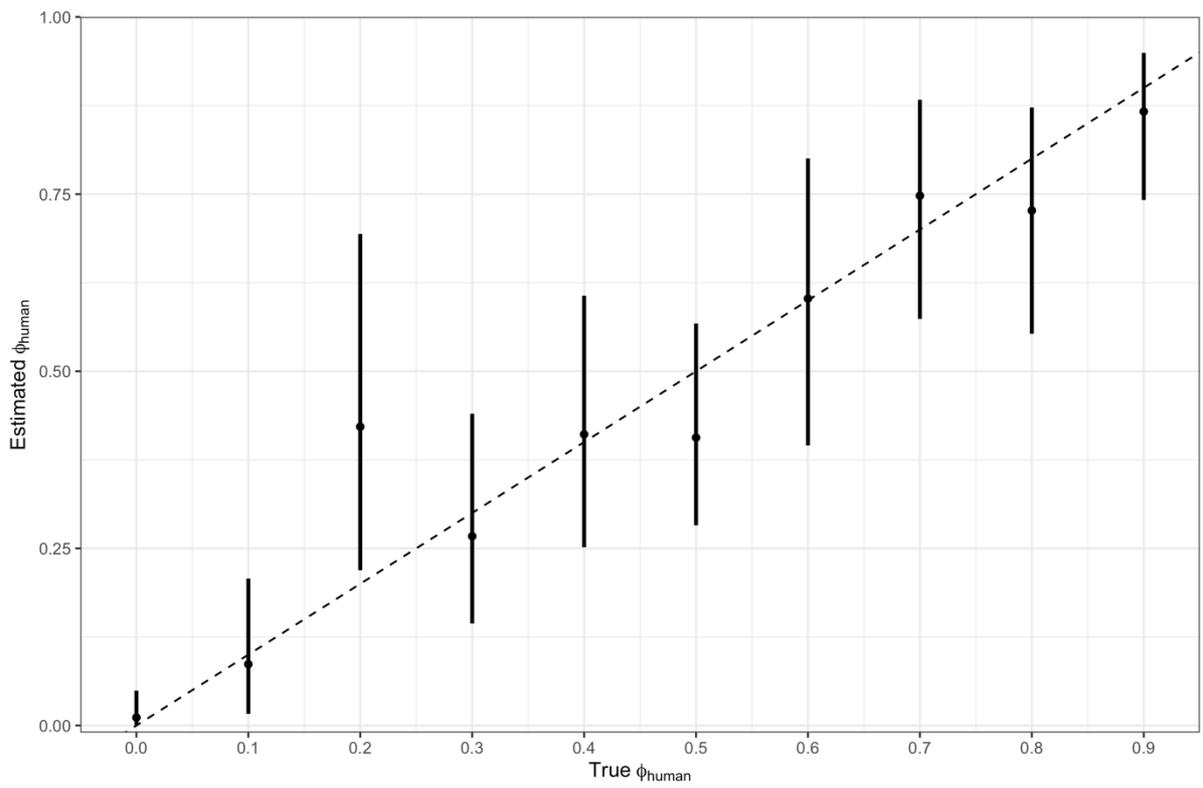

c)

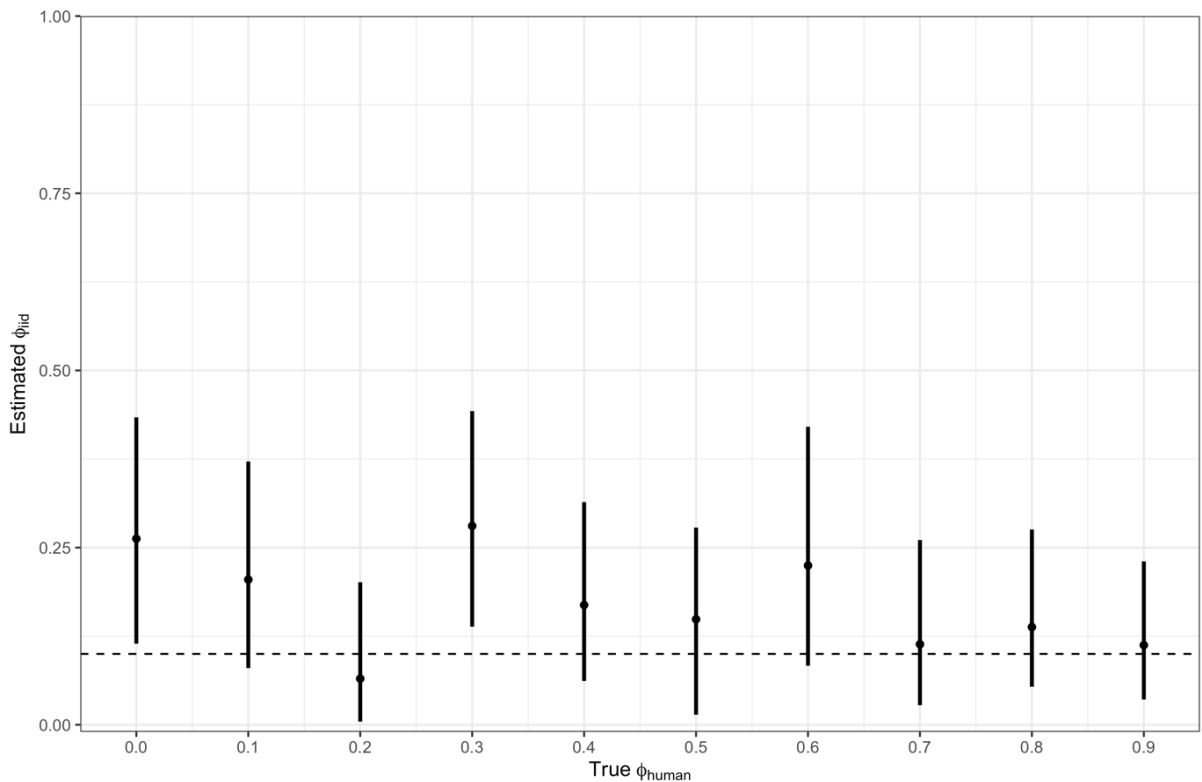

**Figure 4:** Mean and 95% credible interval of the proportion of variance of the random effects explained by a) the distance-based structured term, b) the human movement-based structured term, and c) unstructured random term. Dashed lines represent the true value from simulations.

## 5. Case study

This case study uses the Bayesian spatial smooth models introduced in previous sections to map the spatial patterns of dengue incidence in South Brazil between 2001 and 2020.

### 5.1 Data description

We obtained annual notified dengue cases for each of South Brazil's 1,013 municipalities between 2001 and 2020 from Brazil's Notifiable Diseases Information System (SINAN), freely available via the Health Information Department, DATASUS (https://datasus.saude.gov.br/informacoes-de-saude-tabnet/). To explore the pattern of disease over the whole period, we took the average annual number of cases over the period and rounded this to the nearest whole number. The annual population for each municipality was obtained from the Brazilian Institute of Statistics and Geography (IBGE) via DATASUS (http://tabnet.datasus.gov.br/cgi/deftohtm.exe?ibge/cnv/poptbr.def) over the same period and aggregated in the same way. We used the population divided by 100,000 as an offset to model the dengue incidence rate (DIR), a measure used by the Brazilian Ministry of Health to monitor dengue outbreaks. South Brazil was previously thought to be protected from dengue due to its temperate climate, with winter temperatures too low for the primary vector, *Aedes aegypti*, to breed and transmit the disease. However, recent studies have shown that the northern part of the South region now experiences outbreaks, thought to be due to increasing temperatures (Figure 5, [30]). The data shows a clear distance-based spatial pattern in this region. However, studies of other temperate regions of South America, such as Argentina, have hypothesised that increased outbreaks in cooler regions may be a result of human movement into previously protected cities [7,31]. Data used in this case study are available from https://github.com/sophie-a-lee/spatial_smooth_framework.

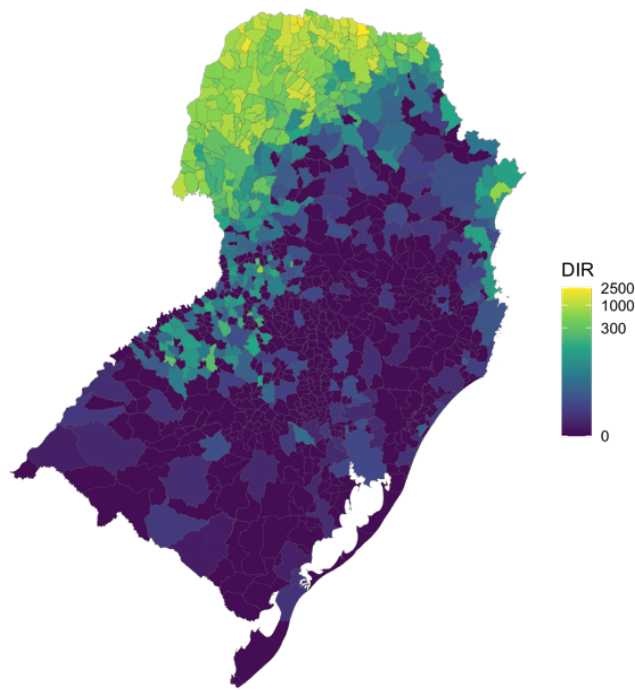

**Figure 5: Average dengue incidence rate (DIR), 2001 - 2020 in South Brazil.**

## 5.2 Modelling approach

We applied a negative binomial model to the average annual dengue cases, using the log of the population divided by 100,000 as an offset to explore the DIR in South Brazil. A negative binomial distribution was assumed to account for possible overdispersion in the dengue case count [5]. Model (12) was applied to the data, spatial random terms were structured by applying thin plate regression splines to the coordinates of the centroids of municipalities ($u_{1,i}$, assuming distance-based connectivity), and human movement-based connectivity coordinates described in Section 4 and the supplementary materials ($u_{2,i}$).

## 5.3 Results

The model found that human movement did not account for much of the spatial structure of the data in this region ($\phi_2$ = 0.003, 95% credible interval (CI): 0, 0.012), and most of the variation could be attributed to the distance-based random term ($\phi_1$ = 0.85, 95% CI: 0.823, 0.876, Figure 6). The human movement data used to create these random effects was only able to capture movement between cities in South Brazil. However, outbreaks in temperate regions such as this are likely triggered by the movement of people from endemic regions elsewhere in Brazil into the South [7].

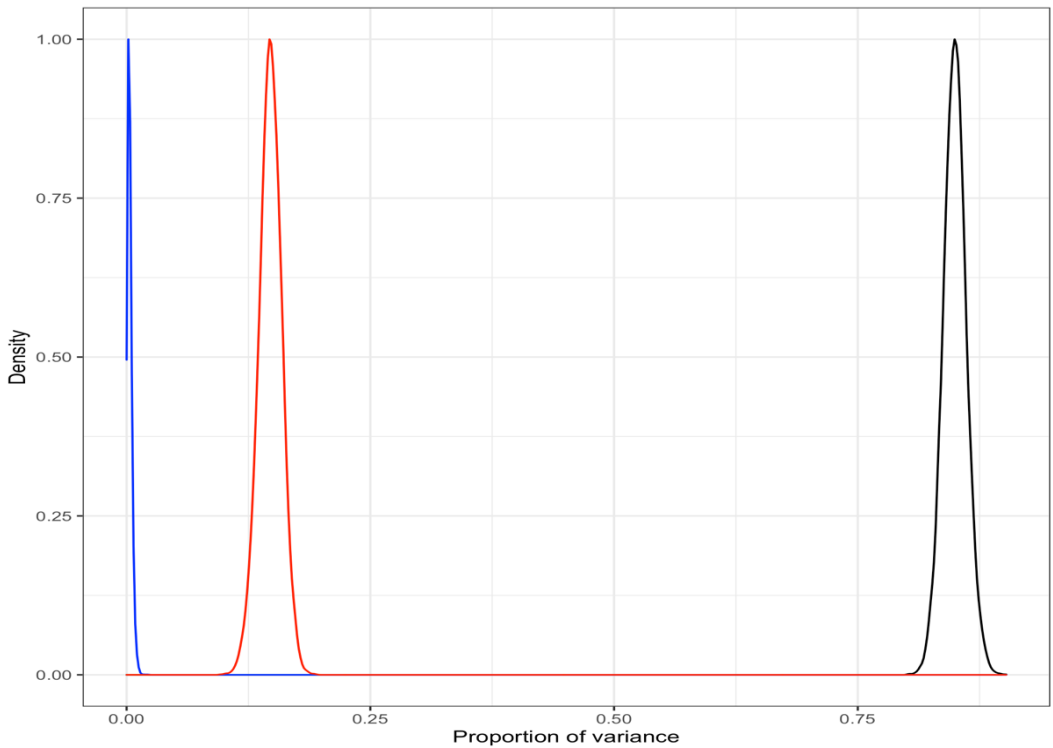

**Figure 6: Estimates of the proportion of variance explained by distance-based (black), human movement-based (blue), and independent (red) random terms.** Using simulations extracted from NIMBLE, the variance of each random term was calculated and divided by the variance of the combined random component, giving the relative contribution of each structure.

Estimates of each random term and the combined total were extracted and plotted to generate hypotheses about these patterns (Figure 7). Most of the spatial structure came from distance-based random term which shows the highest risk was in the north-west, and that the risk decreased to the south. This area of increased risk is the same region which was found to have an increase in the number of months per year with temperatures suitable for dengue transmission since 2010 in a previous study [30]. This model could be extended to include temperature and other variables known to influence dengue risk.

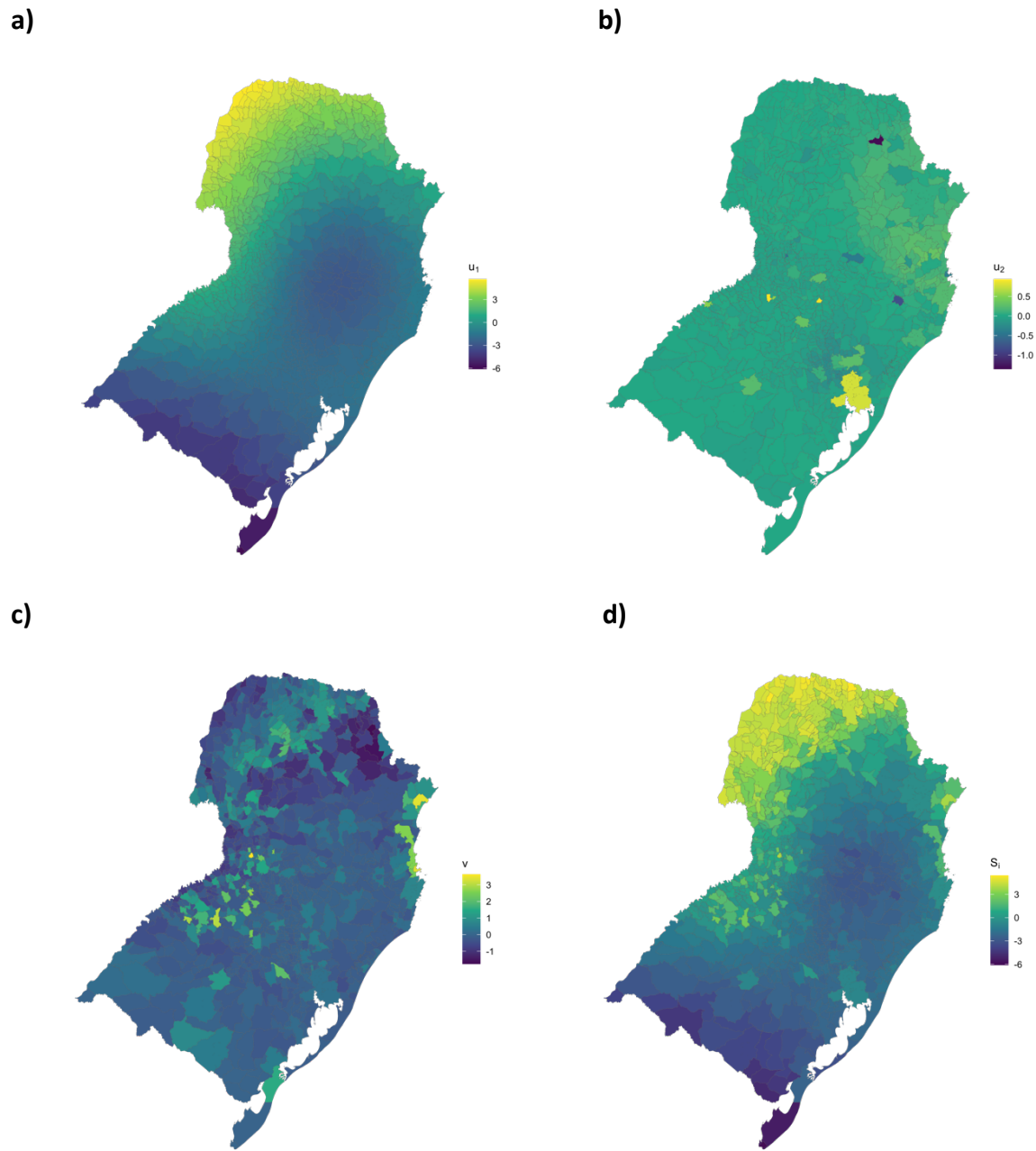

**Figure 7: Mean estimates of the a) distance-based, b) human movement-based, c) unstructured, and d) combined random terms.**

## 6. Discussion

In this paper, we have shown that penalised smoothing splines present a flexible alternative to CAR-based structures of spatial random effects that allow multiple sources of spatial connectivity to be considered within the same model. Smoothing splines allow the spatial structure to be derived from data as part of the model fitting process, producing a non-stationary spatial surface specific to the data being considered. This smooth surface can be extracted and plotted to generate hypotheses about the reasons for this spatial connectivity which may help identify potential drivers of disease. Although many disease mapping studies assume a distance-based structure of connectivity, the smooth spline approach used here can be applied to any continuous measure of connectivity, including human movement. Another benefit of the smoothing spline approach is that the model structure can be extended to include multiple sources of spatial connectivity and can produce parameters quantifying the relative contribution of each structure to the underlying variance of the data. Although this study has focused on disease mapping models of count data, we have shown this method is compatible with other models, such as logistic models for binary data (see the supplementary materials).

Formulating models in NIMBLE (or other similar coding languages) and implementing them using MCMC methods allows for flexibility and complexity in the model structure. However, these models are more likely to face issues with convergence than approximate methods such as INLA [11]. MCMC methods may also take longer than INLA to fit models if convergence is an issue, although this is not always the case when using NIMBLE [32]. One of the main benefits of using penalised smoothing splines over CAR-based priors is that they can be applied to any continuous measure of connectivity. However, the most appropriate measure may not always be clear or available. For example, human movement-based connectivity can be captured using data to describe regular, short-distant movement such as commuting within a city, or long-distance, long-term movement such as migration, which requires different assumptions [33]. Mobile phone data has potential to describe short-term movements at small spatial scales but may be difficult to obtain and care must be taken in some settings where bias may arise [34]. Movement models, such as gravity and radiation models, assume that the number of people moving between areas can be described as a function of population and distance [19]. Movement models provide an alternative when data is unavailable or inappropriate, and have been shown to replicate patterns of movement in large cities and European countries [20,35]. However care must be taken when parameterising these models, particularly in rural settings [36].

The models presented in this study consider a single time point (or data summarised over a given time period), however disease risk is likely to vary over time and models may be required to account for inter-annual or seasonal variation. Data presented in the South Brazil study has been used elsewhere to show the expansion of dengue outbreaks into the region and the changes in spatial structure over the past 20 years [30,37]. The models presented here can be extended to include temporal covariates or random terms to account for seasonal and annual trends. Tensor smooth functions, a type of smoothing spline which allows interaction between variables measured on different scales [24], may be incorporated to explore the interactions between time and connectivity. These structures can be explored to understand changing dynamics of diseases and generate hypotheses about drivers of change or highlight areas at risk. Covariates such as climate indicators can also be included into the models and random term estimates compared to highlight the relative variability in the disease risk explained by these covariates.

Penalised smoothing splines present a flexible alternative to conventional random effect structures when constructing Bayesian hierarchical models. They require minimal user assumptions beyond smoothness and can be applied to any continuous measure of connectivity. By taking a Bayesian view of these smoothing splines, we can incorporate multiple sources of spatial connectivity into a complex modelling framework efficiently and quantify their relative contribution to the overall spatial structure of the data. This is particularly useful in infectious disease epidemiology where the drivers of transmission may be complicated and not fully understood.

## Funding statement

RL was funded by a Royal Society Dorothy Hodgkin Fellowship (grant number DH150120) and a Research Grant for Research Fellows (grant number RGF\R1\180018), which supported a PhD studentship for SAL. TE was funded by the European Union's Horizon 2020 research and innovation programme under grant agreement No. 856612 and the Cyprus Government.

# References

1. Wakefield J. 2007 Disease mapping and spatial regression with count data. *Biostatistics* **8**, 158–183.

2. Lee D. 2011 A comparison of conditional autoregressive models used in Bayesian disease mapping. *Spat. Spatio-Temporal Epidemiol.* **2**, 79–89. (doi:10.1016/j.sste.2011.03.001)

3. Riebler A, Sørbye SH, Simpson D, Rue H. 2016 An intuitive Bayesian spatial model for disease mapping that accounts for scaling. *Stat. Methods Med. Res.* **25**, 1145–1165. (doi:10.1177/0962280216660421)

4. Lee SA, Jarvis CI, Edmunds WJ, Economou T, Lowe R. In press. Spatial connectivity in mosquito-borne disease models: a systematic review of methods and assumptions. *J. R. Soc. Interface* **18**, 20210096. (doi:10.1098/rsif.2021.0096)

5. Lowe R *et al.* 2021 Combined effects of hydrometeorological hazards and urbanisation on dengue risk in Brazil: a spatiotemporal modelling study. *Lancet Planet. Health* **5**, e209–e219. (doi:10.1016/S2542-5196(20)30292-8)

6. Lana RM, Gomes MF da C, de Lima TFM, Honório NA, Codeço CT. 2017 The introduction of dengue follows transportation infrastructure changes in the state of Acre, Brazil: A network-based analysis. *PLoS Negl. Trop. Dis.* **11**. (doi:10.1371/journal.pntd.0006070)

7. Robert MA, Tinunin DT, Benitez EM, Ludueña-Almeida FF, Romero M, Stewart-Ibarra AM, Estallo EL. 2019 Arbovirus emergence in the temperate city of Córdoba, Argentina, 2009–2018. *Sci. Data* **6**, 276. (doi:10.1038/s41597-019-0295-z)

8. Wood SN. 2017 *Generalized additive models: an introduction with R*. CRC press.

9. Wood SN. 2011 Fast stable restricted maximum likelihood and marginal likelihood estimation of semiparametric generalized linear models. *J. R. Stat. Soc. Ser. B Stat. Methodol.* **73**, 3–36. (doi:10.1111/j.1467-9868.2010.00749.x)

10. MacNab YC. 2022 Bayesian disease mapping: Past, present, and future. *Spat. Stat.* , 100593. (doi:10.1016/j.spasta.2022.100593)

11. Rue H, Martino S, Chopin N. 2009 Approximate Bayesian inference for latent Gaussian models by using integrated nested Laplace approximations. *J. R. Stat. Soc. Ser. B Stat. Methodol.* **71**, 319–392. (doi:https://doi.org/10.1111/j.1467-9868.2008.00700.x)

12. Wood S, Wood MS. 2015 Package 'mgcv'. *R Package Version* **1**, 729.

13. Wood SN. 2016 Just Another Gibbs Additive Modeler: Interfacing JAGS and mgcv. *J. Stat. Softw.* **75**, 1–15. (doi:10.18637/jss.v075.i07)

14. Pedersen EJ, Miller DL, Simpson GL, Ross N. 2019 Hierarchical generalized additive models in ecology: an introduction with mgcv. *PeerJ* **7**, e6876.

15. Wood SN. 2004 Stable and efficient multiple smoothing parameter estimation for generalized additive models. *J. Am. Stat. Assoc.* **99**, 673–686.

16. Wood SN. 2003 Thin plate regression splines. *J. R. Stat. Soc. Ser. B Stat. Methodol.* **65**, 95–114.

17. Crainiceanu C, Ruppert D, Wand MP. 2005 Bayesian analysis for penalized spline regression using WinBUGS.

18. Cox MA, Cox TF. 2008 Multidimensional scaling. In *Handbook of data visualization*, pp. 315–347. Springer.

19. Simini F, González MC, Maritan A, Barabási A-L. 2012 A universal model for mobility and migration patterns. *Nature* **484**, 96–100.

20. Tizzoni M, Bajardi P, Decuyper A, Kon Kam King G, Schneider CM, Blondel V, Smoreda Z, González MC, Colizza V. 2014 On the use of human mobility proxies for modeling epidemics. *PLoS Comput. Biol.* **10**, e1003716. (doi:10.1371/journal.pcbi.1003716)

21. De Valpine P *et al.* 2021 Nimble: MCMC, particle filtering, and programmable hierarchical modeling. *R Package Version 011* **1**.

22. de Valpine P, Turek D, Paciorek CJ, Anderson-Bergman C, Lang DT, Bodik R. 2017 Programming with models: writing statistical algorithms for general model structures with NIMBLE. *J. Comput. Graph. Stat.* **26**, 403–413.

23. R Core Team. 2021 *R: A language and environment for statistical computing*. 4.1.1. Vienna, Austria. See https://www.R-project.org/.

24. Wood SN. 2006 Low-Rank Scale-Invariant Tensor Product Smooths for Generalized Additive Mixed Models. *Biometrics* **62**, 1025–1036. (doi:10.1111/j.1541-0420.2006.00574.x)

25. Simpson D, Rue H, Riebler AI, Martins TG, Sørbye SH. 2017 Penalising Model Component Complexity: A Principled, Practical Approach to Constructing Priors. (doi:10.1214/16-STS576)

26. Rue H, Riebler A, Sørbye SH, Illian JB, Simpson DP, Lindgren FK. 2017 Bayesian computing with INLA: a review. *Annu. Rev. Stat. Its Appl.* **4**, 395–421.

27. Besag J. 1974 Spatial interaction and the statistical analysis of lattice systems. *J. R. Stat. Soc. Ser. B Methodol.* **36**, 192–225.

28. Martino S, Rue H. 2009 Implementing approximate Bayesian inference using Integrated Nested Laplace Approximation: A manual for the inla program. *Dep. Math. Sci. NTNU Nor.*

29. Gelman A, Hwang J, Vehtari A. 2014 Understanding predictive information criteria for Bayesian models. *Stat. Comput.* **24**, 997–1016.

30. Lee SA, Economou T, Catão R de C, Barcellos C, Lowe R. 2021 The impact of climate suitability, urbanisation, and connectivity on the expansion of dengue in 21st century Brazil. *PLoS Negl. Trop. Dis.* **15**, e0009773. (doi:10.1371/journal.pntd.0009773)

31. Robert MA, Stewart-Ibarra AM, Estallo EL. 2020 Climate change and viral emergence: evidence from Aedes-borne arboviruses. *Curr. Opin. Virol.* **40**, 41–47. (doi:10.1016/j.coviro.2020.05.001)

32. Lawson AB. 2021 *Using R for Bayesian Spatial and Spatio-temporal Health Modeling*. CRC Press.

33. Stoddard ST, Morrison AC, Vazquez-Prokopec GM, Soldan VP, Kochel TJ, Kitron U, Elder JP, Scott TW. 2009 The role of human movement in the transmission of vector-borne pathogens. *PLoS Negl Trop Dis* **3**, e481.

34. Wesolowski A, Buckee CO, Engø-Monsen K, Metcalf CJE. 2016 Connecting mobility to infectious diseases: the promise and limits of mobile phone data. *J. Infect. Dis.* **214**, S414–S420.

35. Kraemer MUG *et al.* 2018 Inferences about spatiotemporal variation in dengue virus transmission are sensitive to assumptions about human mobility: a case study using geolocated tweets from Lahore, Pakistan. *Epj Data Sci.* **7**, 17. (doi:10.1140/epjds/s13688-018-0144-x)

36. Meredith HR *et al.* 2021 Characterizing human mobility patterns in rural settings of sub-Saharan Africa. *eLife* **10**, e68441. (doi:10.7554/eLife.68441)

37. Codeco CT *et al.* 2022 Fast expansion of dengue in Brazil. *Lancet Reg. Health - Am.* **12**, 100274. (doi:10.1016/j.lana.2022.100274)

# Supplementary material

## S1 Human movement coordinates

To create a smooth surface describing the spatial structure of data under a given connectivity assumption, we apply penalised smoothing splines to coordinates describing this relative connectivity. For example, when describing distance-based connectivity, we apply smoothing splines to the coordinates of observations. However, when a coordinate system does not currently exist that describes the relative connectivity between observations, we must create one. Multidimensional scaling (MDS) is a mathematical approach that translates continuous measures of distance (in this case, connectivity) onto an abstract cartesian space and returns a set of coordinates [1]. When considering human movement-based connectivity, we can apply MDS to a continuous measure of human movement, i.e., the number of people moving between areas, and use the resulting coordinates to construct the smooth spatial surface.

Continuous measures of human movement-based connectivity can include observed data, such as the number of air travel passengers, or can be estimated using movement models, such as gravity and radiation models [2,3], that assume the number of people moving between areas is a function of population and distance. To imitate these movement models, we used smoothing splines to investigate the relationship between 'connectedness', distance and population. In Brazil, the Regiões de Influência das Cidades [Regions of Influence of Cities] (REGIC) study aims to recreate the Brazilian urban network which explains the movement of people, goods, and services around the country [4]. Based on a survey of residents, the study produced a binary matrix classifying all cities within Brazil as either 'connected' or not in 2018 (Figure M1). We used a logistic generalised additive model to estimate the relationship between the binary measure of connectedness, and the Euclidean distance and population in each city:

$$c_{i,j} \sim Bernoulli(p_{i,j})$$

$$logit(p_{i,j}) = \alpha + f(d_{i,j}, r_i, r_j)$$

Where $p_{i,j}$ is the probability that cities $i$ and $j$ are connected, $c_{i,j}$ was the binary connectedness indicator taken from the REGIC study, $f(d_{i,j}, r_i, r_j)$ is a 3-dimensional tensor smoothing spline applied to the Euclidean distance between cities ($d_{i,j}$) and the populations in city $i$ ($r_i$) and city $j$ ($r_j$). Tensor smoothing splines allow interactions between covariates measured on different scales (in this example, population and distance) [5]. The model was implemented using the mgcv package [6] using restricted maximum likelihood (REML), allowing for Bayesian interpretation of the results [7,8]. The predicted probability that cities are connected ($p_{i,j}$) was extracted from the model by taking simulations from the posterior distribution. MSD was applied to the predicted probabilities to produce a coordinate system describing the relative connectivity of municipalities based on human movement.

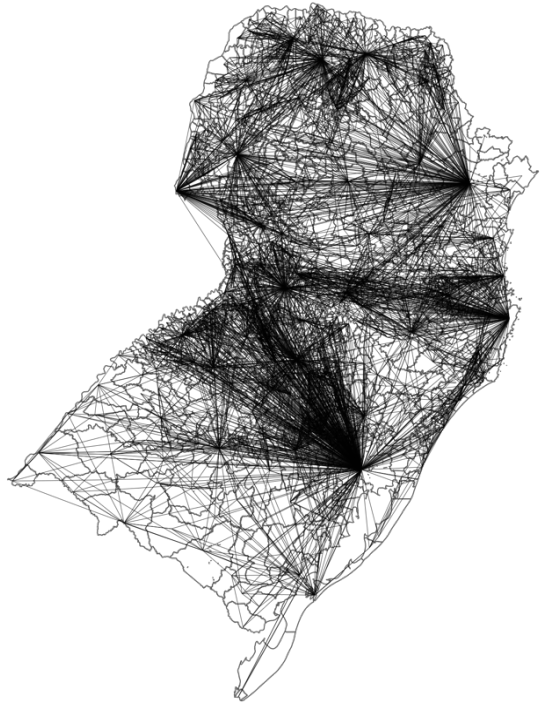

**Figure M1: Connections between municipalities in South Brazil extracted from the REGIC 2018 study [4].**

# S2 Simulation study: a single source of human movement-based spatial structure

In this section, we present a simulation study with a single source of spatial connectivity in the data, arising due to human movement.

## S2.1 Data generation

Fictitious disease data was generated from a Poisson distribution for each of the 1,013 municipalities in South Brazil:

$$y_i \sim Poisson(E(y_i))$$

$$log(E(y_i)) = log(\xi_i) + \alpha + S_i \quad (1)$$

Where $y_i$ is the number of cases in municipality $i$,, $E(y_i)$ is the expected count, and $\xi_i$ is an offset term set to the population divided by 100,000 so that $y_i/\xi_i$ is the incidence rate per 100,000 residents for municipality $i$. $\beta = 0$ is the intercept, or baseline risk, and $S_i$ is the underlying spatial structure of the data:

$$S_i = \sqrt{\phi} \cdot sm(x_i, z_i) + \sqrt{(1-\phi)} \cdot \varepsilon_i \quad (2)$$

Where $\phi$ is a mixing parameter that determines the relative contribution of a spatially structured surface, $sm(x_i, z_i)$, and an unstructured random term, $\varepsilon_i \sim N(0,1)$. $sm(x_i, z_i)$ is a smooth function applied to human movement-based connectivity coordinates, defined above, which emulates a spatially structured surface (taken from [5]):

$$sm(x, z) = \pi \sigma_x \sigma_z \left( 1.2 e^{-(x-0.2)^2/\sigma_x^2 - (z-0.3)^2/\sigma_z^2} + 0.8 e^{\frac{-(x-0.7)^2}{\sigma_x^2} - \frac{(z-0.8)^2}{\sigma_z^2}} \right) \quad (3)$$

$$\sigma_x = 0.3, \quad \sigma_z = 0.4$$

The smooth function was centred around 0 by subtracting the overall mean from each value. 11 simulated datasets were produced using equation (1), setting values of $\phi$ between 0 and 1 at intervals of 0.1 (Figure M2).

a)

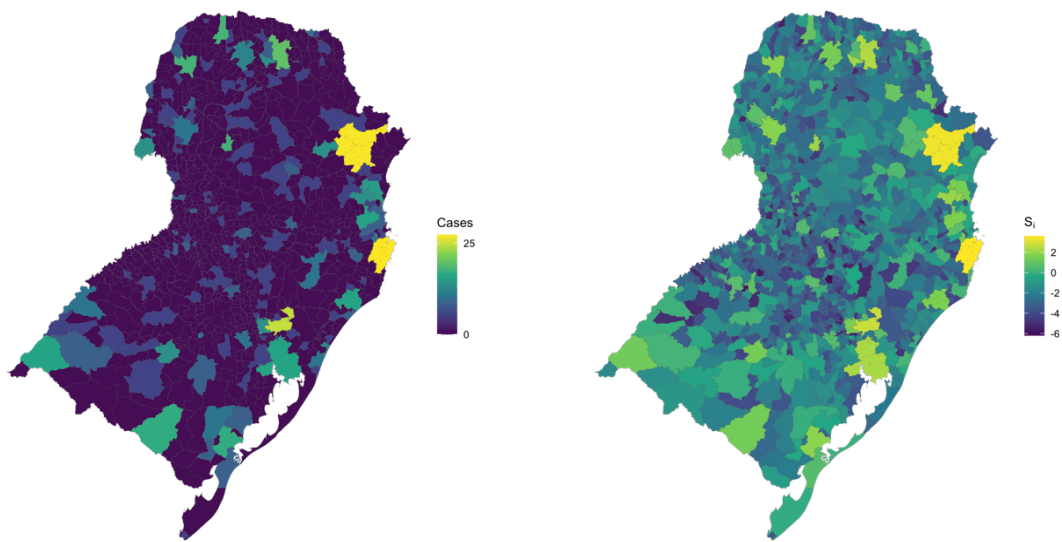

b)

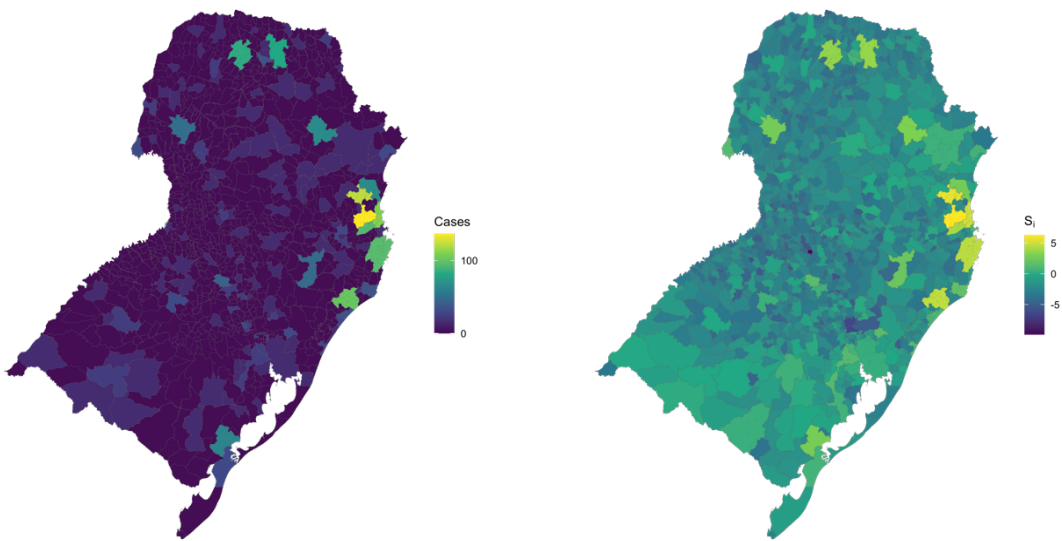

c)

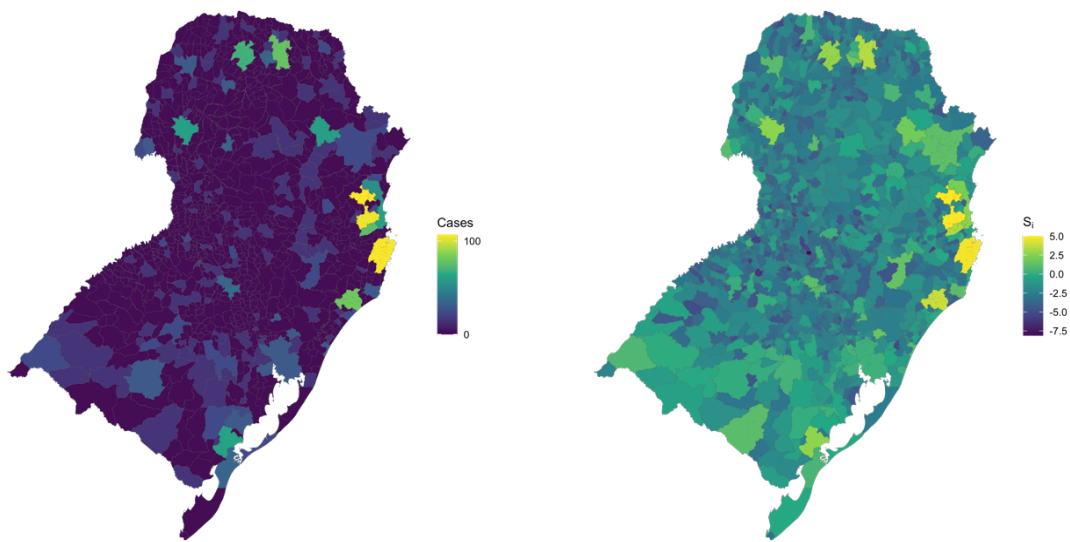

**Figure M2: Simulated disease counts (left) and spatial random effects (right) under a human movement-based structure using different spatial structure combinations.** The number of cases simulated from a Poisson model and the underlying spatial structure where the data has a) no spatial structure ($\phi = 0$), b) a human movement-based structure only ($\phi = 1$), and c) equal contribution of both structures ($\phi = 0.5$).

### S2.2 Modelling approach

A spatial Poisson model containing spatially structured and unstructured random effects was applied to each simulated dataset:

$$y_i \sim Poisson(E(y_i))$$

$$log(E(y_i)) = log(\xi_i) + \alpha + u_i + v_i \quad (4)$$

$u_i$ is a spatially structured random term, created by applying a thin plate regression spline to human movement-based coordinates generated using MSD (see section S1). $v_i$ is a spatially unstructured term, expected to follow a zero-mean normal distribution, representing heterogeneity between regions. The smooth surface used to structure $u_i$ was generated using the mgcv package and extracted using the jagam function [6,9]. Models were implemented using MCMC via NIMBLE [10,11]. The relative contribution of each random term to the overall marginal variance was calculated from simulations and compared to the known value of $\phi$ used to generate the data (see equation (2)).

### S2.3 Results

The 95% credible interval of the estimated intercept contained the true value of zero for all but one of the simulations (Figure M3). The proportion of the random effect variance explained by the structured term provided an accurate estimate of the known contribution from each simulation (Figure M4), showing that alternative spatial structures can be considered within the proposed modelling framework. Multidimensional scaling allows coordinate systems to be derived from any continuous measure of connectivity, allowing assumptions of connectivity beyond distance-based to be included and tested.

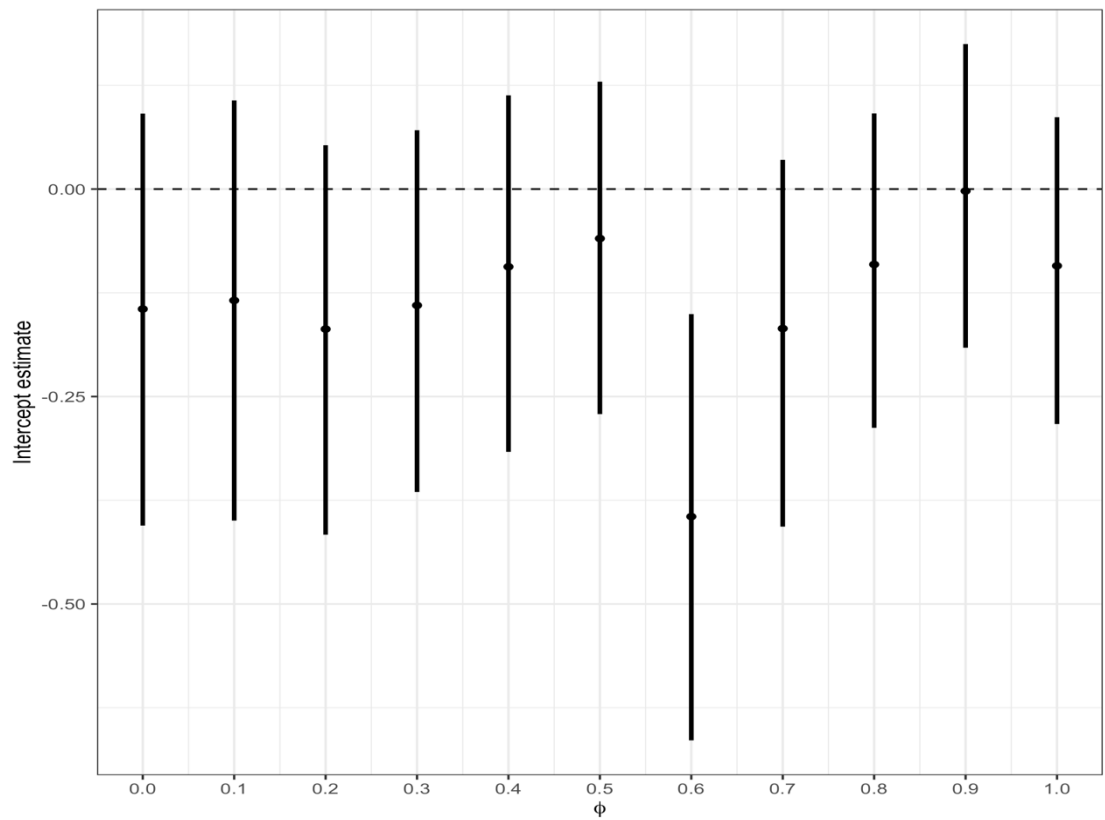

**Figure M3: Mean and 95% credible interval of intercept coefficient estimates compared to the true simulated value, 0.**

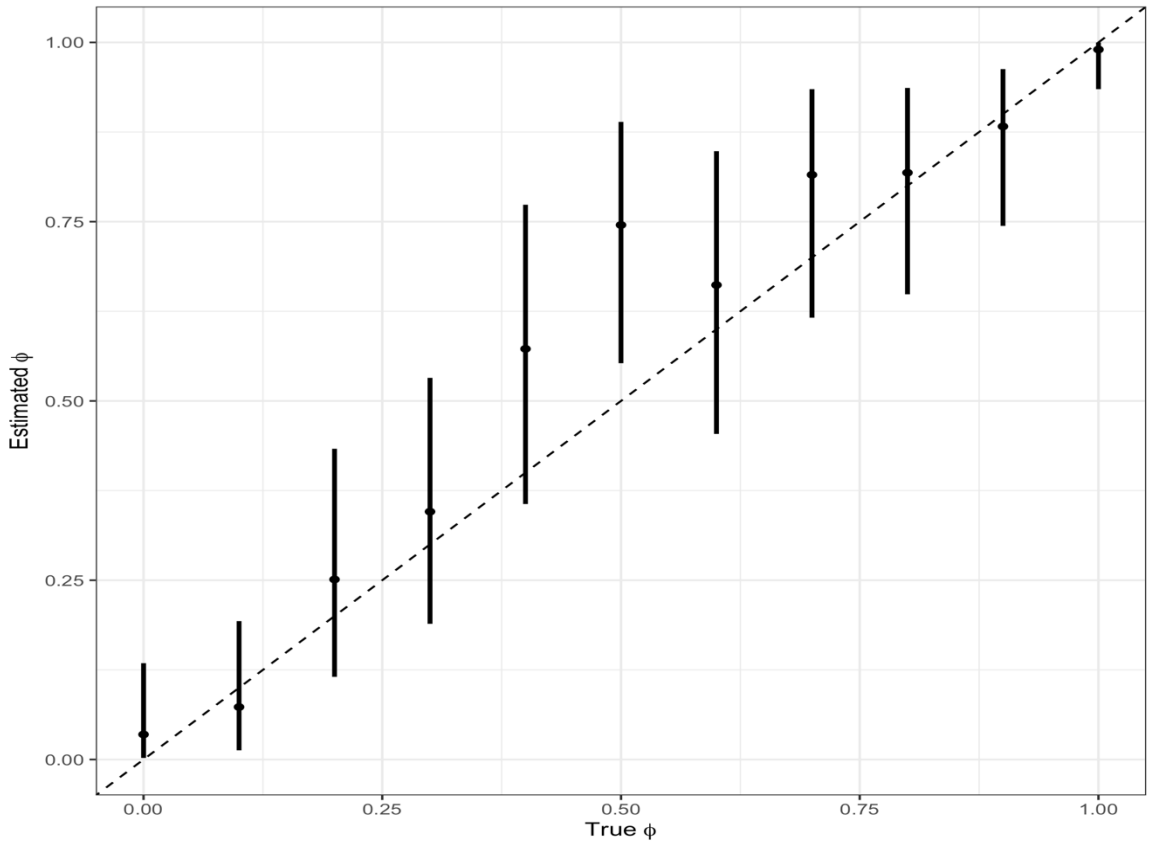

**Figure M4: The mean and 95% credible interval of estimated $\phi$ values compared to the known value.** Estimated $\phi$ values are calculated using the proportion of the random effect variance explained by the spatially structured term.

# S3 Simulation study: spatial modelling of binary data

Although the primary purpose of this paper is to develop a modelling framework compatible with count data, the smoothing spline approach could be used to structure random terms for other models, such as logistic models for binary outcomes. In this simulation study, we compare the proposed penalised smoothing spline approach to INLA's BYM2 model applied to a generated binary response.

## S3.1 Data generation

Binary response data, $y_i$, was generated for each of South Brazil's 1,013 municipalities from a binomial distribution with 20 trials or events, with a distance-based spatial structure:

$$y_i \sim Binomial(p_i, 20),$$

$$log\left(\frac{p_i}{1-p_i}\right) = \beta + S_i \quad (5)$$

Where $\beta$ is an intercept set to 0, $p_i$ is the probability of an event, and $S_i$ is the spatial structure of the data, defined as:

$$S_i = \sqrt{\phi} \cdot sm(x_i, z_i) + \sqrt{(1-\phi)} \cdot \varepsilon_i$$

Where $\phi$ is a mixing parameter defining the relative contribution of a spatially structured term, $sm(x_i, z_i)$, and an unstructured term, $\varepsilon_i \sim N(0,1)$. A smooth function, $sm$ (see equation (3)), was applied to the coordinates of the centroid of municipalities that were scaled to take values between 0 and 1. $sm(x_i, z_i)$ was centred around 0 by subtracting the overall mean from each value. 11 simulated datasets were produced by setting $\phi$ values between 0 and 1 at intervals of 0.1 (Figure M5).

a)

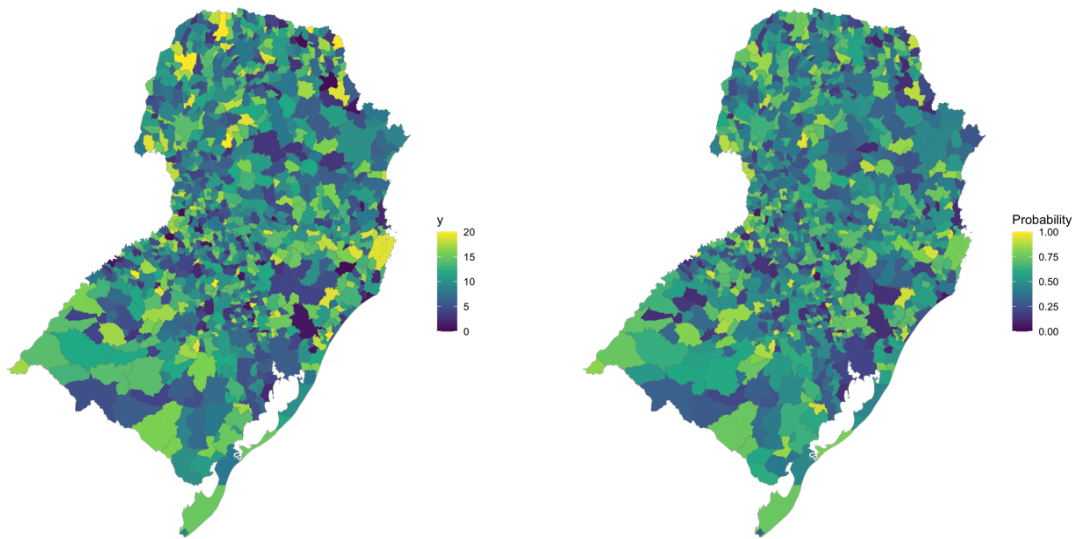

b)

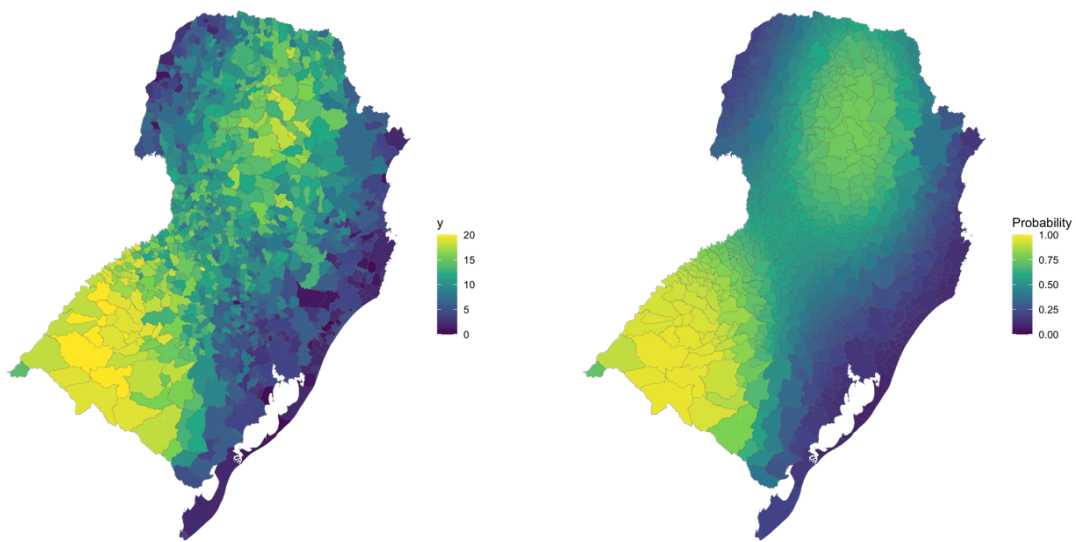

c)

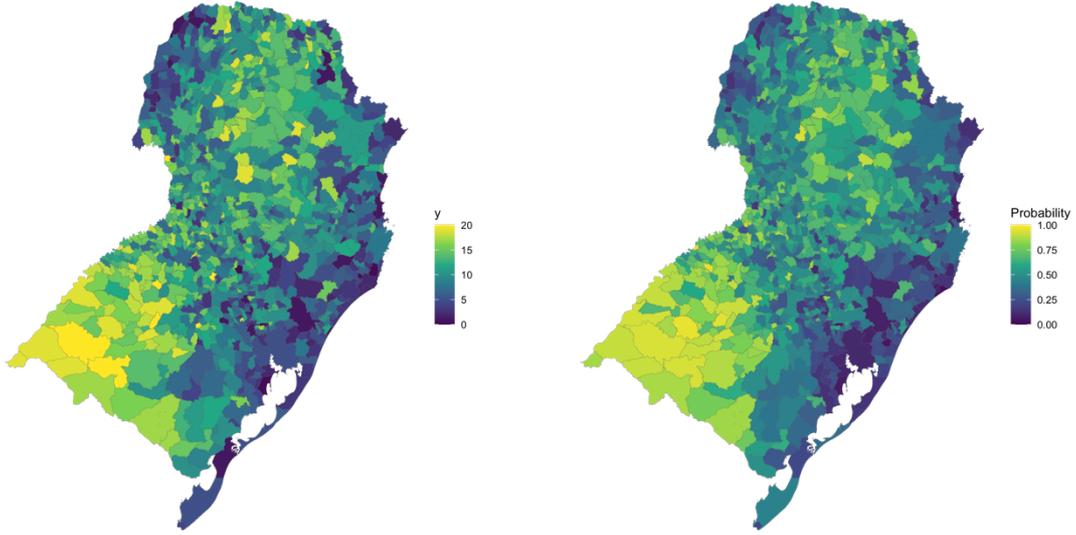

**Figure M5: Simulated data (left) and probability of an event (right) under a distance-based structure using different spatial structure combinations.** The number of events simulated from a binomial model and the probability of an event where the data has a) no spatial structure ($\phi = 0$), b) a distance-based structure only ($\phi = 1$), and c) equal contribution of both structures ($\phi = 0.5$).

## S3.2 Modelling approach

Two logistic models containing spatially structured and unstructured random effects were applied to each simulated dataset:

$$y_i \sim Binomial(p_i, 20)$$

$$log\left(\frac{p_i}{1-p_i}\right) = \alpha + u_i + v_i \quad (6)$$

$$log\left(\frac{p_i}{1-p_i}\right) = \alpha + \frac{1}{\tau}(\sqrt{\phi}u_{*i} + \sqrt{1-\phi}v_{*i}) \quad (7)$$

In model (6), $u_i$ is a spatially structured random term representing spatial connectivity within the data, created be applying a thin plate regression spline to coordinates of the centroid of each municipality. $v_i$ is a spatially unstructured term, expected to follow a zero-mean normal distribution, representing heterogeneity between regions. This spatial smooth model was compared to a CAR-based random effect model using R-INLA's BYM2 model specification, given in equation (7) [12–14]. Here, $u_{*i}$ is a spatially structured random effect assuming a CAR structure, where municipalities are considered connected if and only if they share a border. $v_{*i}$ are unstructured random effects, and $\phi$ are mixing parameters that measure the relative contribution of the structured and unstructured random effects to the overall marginal variance $\left(\frac{1}{\tau^2}\right)$ of the random effect. The penalised smoothing spline structure was generated using R's mgcv package [15] and extracted using the jagam function [9]. Spatial smooth models were implemented using Markov chain Monte-Carlo (MCMC) simulations in R via the NIMBLE package [10]. The CAR model was created using INLA's BYM2 model [13,16].

We compared model performance between the spatial smooth and BYM2 models using receiver operating characteristic (ROC) curves, a comparison of the true positive and true negative rates, the Brier score, the mean squared difference between the predicted probability and observed outcomes, and WAIC. Higher values of area under the ROC (AUROC) curve and lower values of the Brier score and WAIC indicate a better model fit. The proportion of the random variance attributed to the spatially structured term was calculated from the smoothing spline model and compared to estimates of $\phi$ extracted from the INLA model output and the known values for each simulation.

### S3.3 Results

Model comparison statistics AUROC were almost equivalent between the spatial smooth and BYM2 models, while the Brier scores and WAIC values preferred the spatial smooth models (Table S1), indicating that the spatial smooth model performs as well (if not better) than one of the current standard approaches. This shows that the smoothing spline structure provides an alternative to CAR-based structures in binomial models as well as models for count data, particularly when the full spatial structure may not be fully understood or where the structure is neither stationary nor isotropic. Both models were able to accurately estimate the intercept coefficient (Figure M6).

**Table S1: Model comparison statistics and mean estimates of the mixing parameter, $\phi$, from the smoothing spline and INLA BYM2 models.** Area under the receiver operating characteristic curve (AUROC), Brier score, WAIC, and $\phi$ estimates extracted from logistic models fitted to simulated data using smoothing splines or INLA's BYM2 model to structure random effects. The optimal goodness-of-fit statistic is given in bold, that is, the highest AUROC, the lowest Brier score and WAIC, and the $\phi$ estimate closest to the value used in each simulation.

|  | Smoothing spline model | | | | INLA BYM2 model | | | |
| --- | --- | --- | --- | --- | --- | --- | --- | --- |
| $\phi$ | AUROC | Brier score | WAIC | $\phi$ estimate | AUROC | Brier score | WAIC | $\phi$ estimate |
| 0 | 0.764 | **0.199** | **4937.1** | **0.008** | 0.764 | 0.246 | 5084.4 | 0.014 |
| 0.1 | 0.764 | **0.199** | **4932.8** | 0.077 | 0.764 | 0.247 | 5075.3 | 0.126 |
| 0.2 | 0.762 | **0.2** | **4926.1** | 0.204 | 0.762 | 0.246 | 5053.6 | 0.283 |
| 0.3 | 0.763 | **0.199** | **4909.4** | 0.278 | 0.764 | 0.246 | 5031.1 | 0.358 |
| 0.4 | 0.759 | **0.201** | **4906.1** | 0.373 | 0.759 | 0.246 | 5002.8 | 0.457 |
| 0.5 | 0.758 | **0.201** | **4884.1** | 0.487 | 0.758 | 0.247 | 4967.6 | 0.533 |
| 0.6 | 0.749 | **0.204** | **4849.7** | 0.615 | 0.749 | 0.247 | 4906.3 | 0.682 |
| 0.7 | 0.749 | **0.204** | **4806.5** | 0.71 | 0.75 | 0.248 | 4857.2 | 0.747 |
| 0.8 | 0.754 | **0.202** | **4733.2** | 0.797 | 0.754 | 0.245 | 4756.5 | 0.934 |
| 0.9 | 0.751 | **0.203** | **4645.8** | 0.877 | 0.754 | 0.246 | 4626.3 | 0.976 |
| 1 | 0.733 | **0.208** | **4289.2** | 0.995 | 0.748 | 0.246 | 4324.6 | **0.998** |

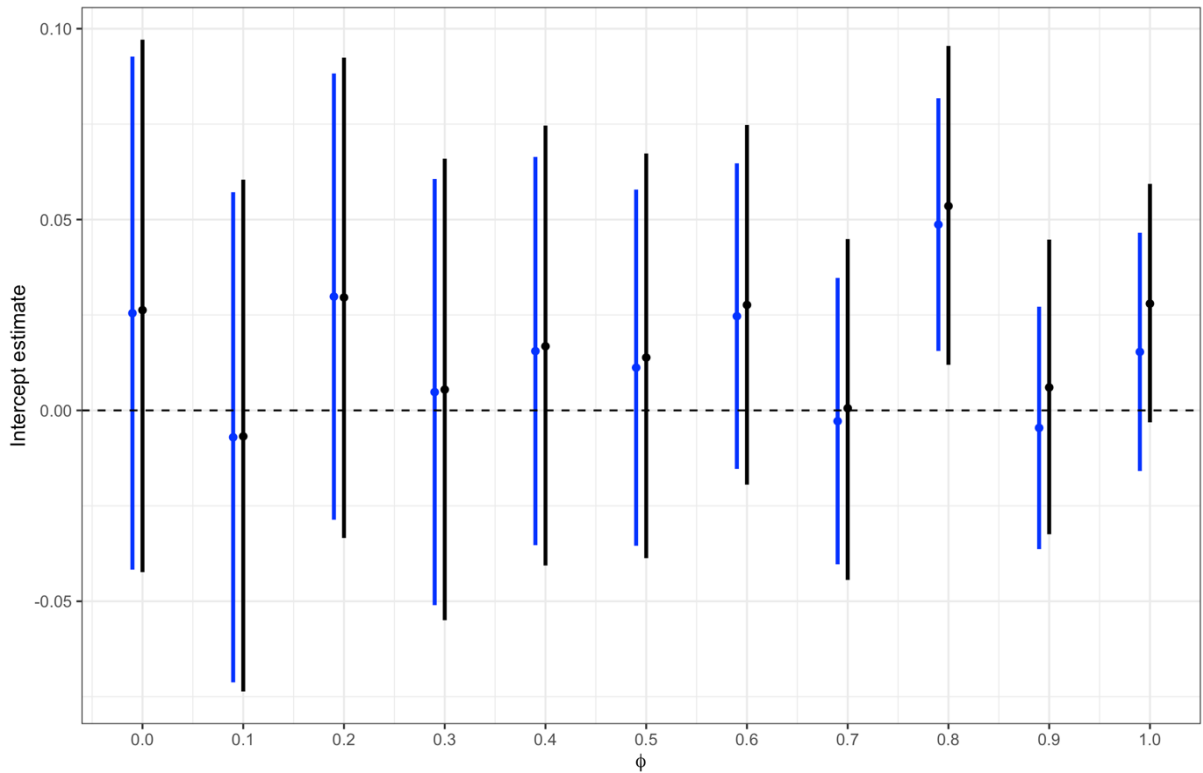

**Figure M6: Mean and 95% credible interval of the intercept coefficient estimates from the smoothing spline (black) and BYM2 (blue) models, compared to the true simulated value, 0.**

The proportion of the random effect variance explained by the spatially structured term in the smoothing spline model provided an accurate estimation of the true value from simulations (Figure M7). This value was comparable to the $\phi$ hyperparameter extracted from the BYM2 model using INLA and can therefore be interpreted in a similar way.

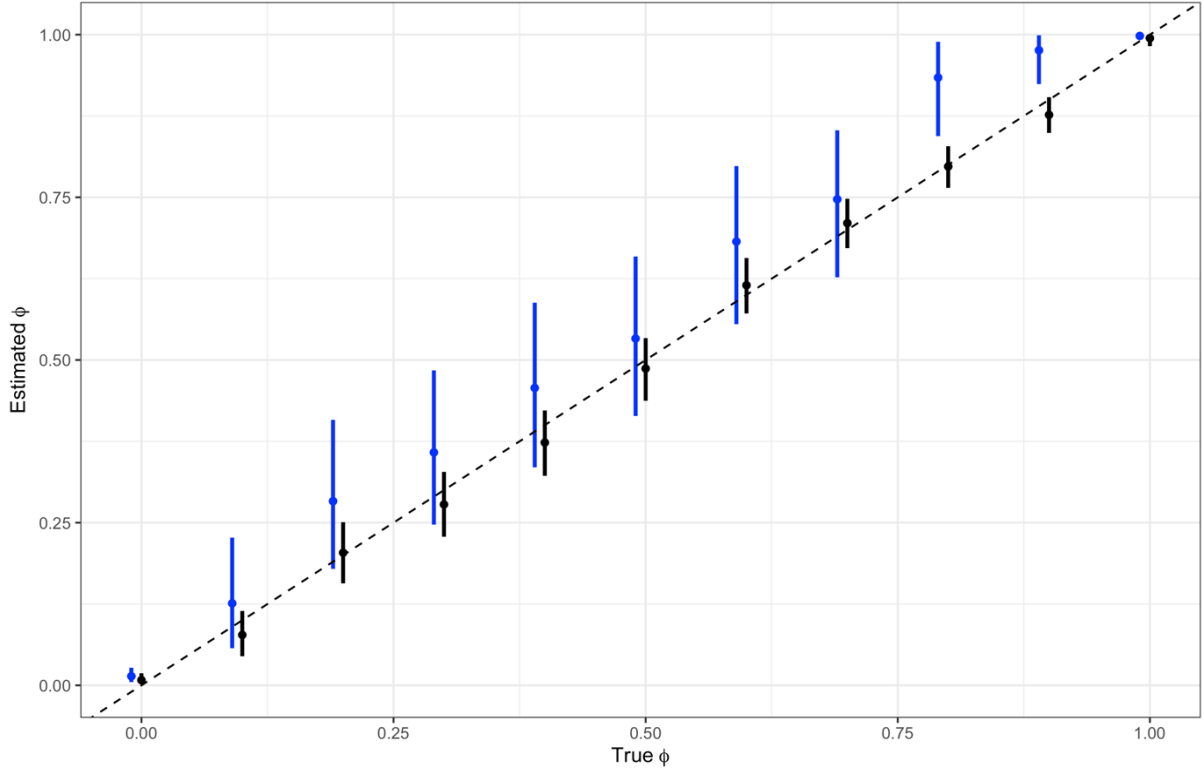

**Figure M7: The mean and 95% credible interval of estimated $\phi$ values extracted from the smoothing spline (black) and INLA BYM2 (blue) models compared to the known value.** Estimated $\phi$ values for the smoothing spline model were calculated using the proportion of the random effect variance explained by the spatially structured term and were extracted from INLA output for the BYM2 model.

## S4 Simulation study: binary data with two sources of spatial structure
The binomial simulation study above was extended to also include a source of human movement-based connectivity.

### S4.1 Data generation
Binary response data, $y_i$, was generated for each of South Brazil's 1,013 municipalities from a binomial distribution with 20 trials or events, using the same process as simulation study in Section S2, but with an extended spatial structure:

$$S_i = \sqrt{\phi_1} \cdot sm(a_i, b_i) + \sqrt{\phi_2} \cdot sm(c_i, d_i) + \sqrt{\phi_3} \cdot \varepsilon_i \quad (8)$$

$$\phi_1 + \phi_2 + \phi_3 = 1$$

$sm$ is a smoothing function (equation (3)), applied to scaled coordinates of the centroid of municipalities, $a_i$ and $b_i$, to create a distance-based smooth term, and applied to human movement-based connectivity coordinates, $c_i$ and $d_i$, derived using multidimensional scaling and the Brazilian REGIC study [4] as described previously. $\varepsilon_i$ is a random draw from a normal distribution ($\varepsilon_i \sim N(0,1)$), representing heterogeneity between municipalities. $\phi_1$, $\phi_2$ and $\phi_3$ are mixing parameters which define the relative contribution of the spatially structured and unstructured terms to the underlying spatial surface. $\phi_3$ was fixed at 0.2, $\phi_1$ and $\phi_2$ took values between 0 and 0.8 at intervals of 0.1 to create 9 simulated datasets (Figure M8).

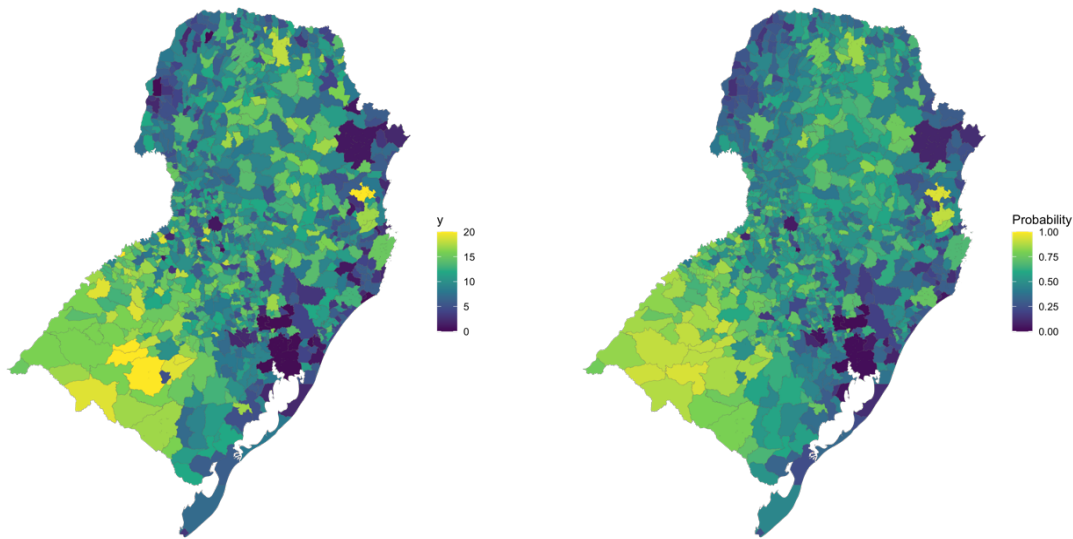

**Figure M8: Simulated data (left) and probability of an event (right) containing two sources of spatial structure.** An example of simulated data used to test the model, where $\phi_1 = 0.4$, $\phi_2 = 0.4$, and $\phi_3 = 0.2$.

### S4.2 Modelling approach

A logistic model containing two spatially structured random terms and one unstructured random terms was applied to each simulated dataset:

$$y_i \sim Binomial(p_i, 20)$$

$$log\left(\frac{p_i}{1-p_i}\right) = \beta + u_{1,i} + u_{2,i} + v_i \quad (9)$$

Where $u_{1,i}$ is a distance-based spatially structured random term, created by applying a thin plate regression spline to the coordinates of the centroid of municipalities. $u_{2,i}$ is a human movement-based spatially structured random term, created by applying a thin plate regression spline to coordinates describing relative connectivity between municipalities based on human movement (see Section S1). $v_i$ is a spatially unstructured term, expected to follow a zero-mean normal distribution, representing heterogeneity between municipalities. Spatially structured terms were generated using R's mgcv package [15] and extracted using the jagam function [9]. Spatial smooth models were fitted using Markov chain Monte-Carlo (MCMC) simulations in R via the NIMBLE package [10].

### S4.3 Results

The extended spatial model was able to accurately estimate the intercept coefficient for each simulated dataset (Figure M9). The ability for each model to detect the contribution of each spatial structure to the random effects varied across simulated datasets and between assumptions of connectivity (Figure M10). Although the model estimates were able to detect increases in the contribution of these structures, the estimates and 95% credible intervals did not always contain the true mixing parameter values from the simulation. In particular, this model underestimated the contribution of human movement to the spatial structure and attributed this to either distance-based or independent, unstructured terms for larger values of $\phi_2$ (Figure M10). Therefore, as with the INLA BYM2 models compared previously, care should be taken when interpreting the estimates of the mixing parameters.

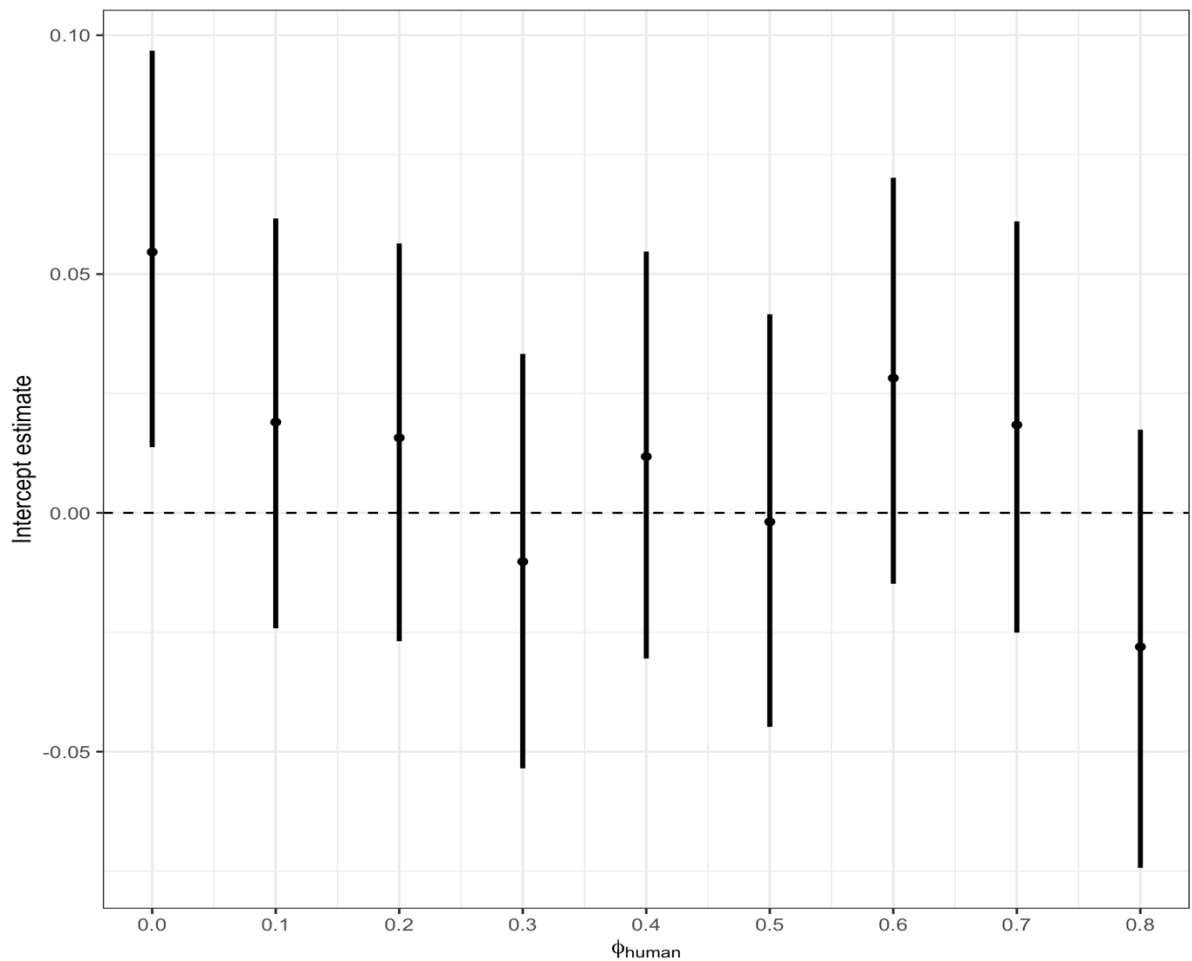

**Figure M9: Mean and 95% credible interval of the intercept coefficient estimates, compared to the true simulated value, 0 (sorted by $\phi_2$).**

a)

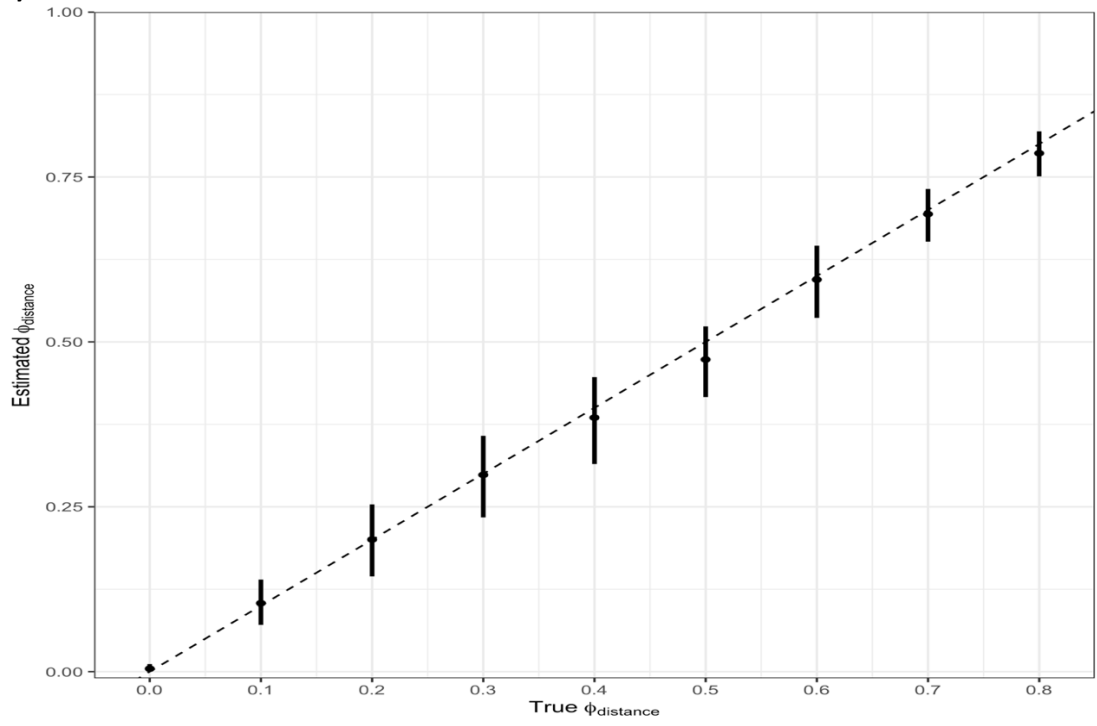

b)

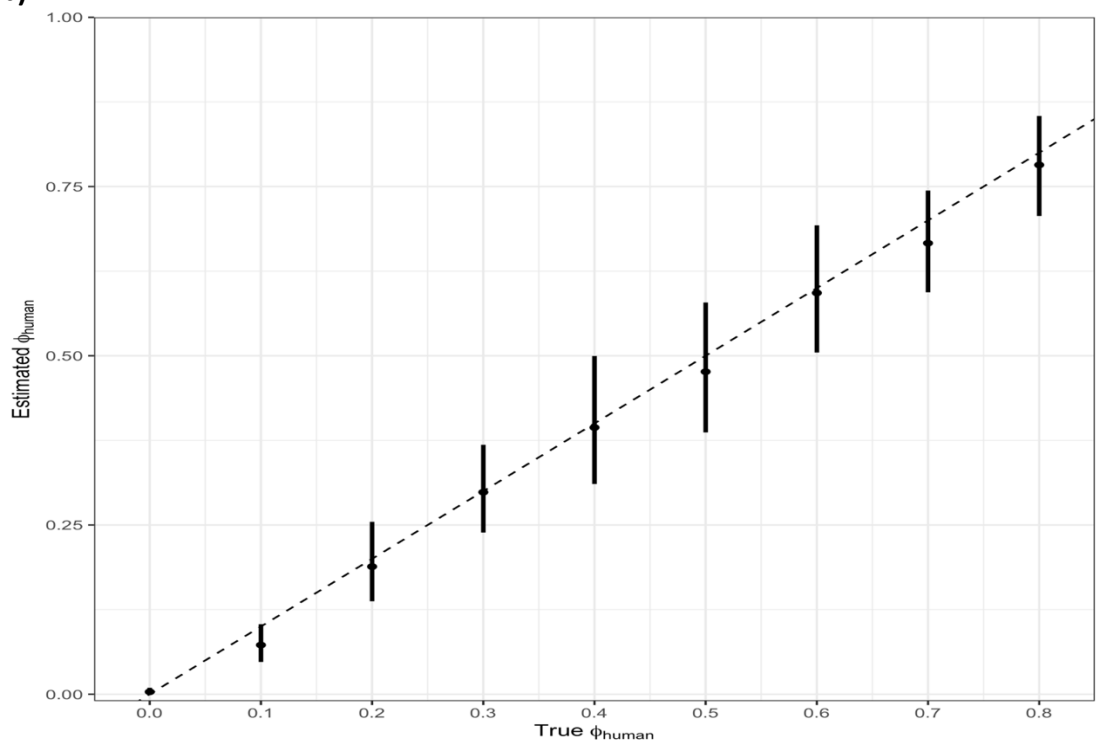

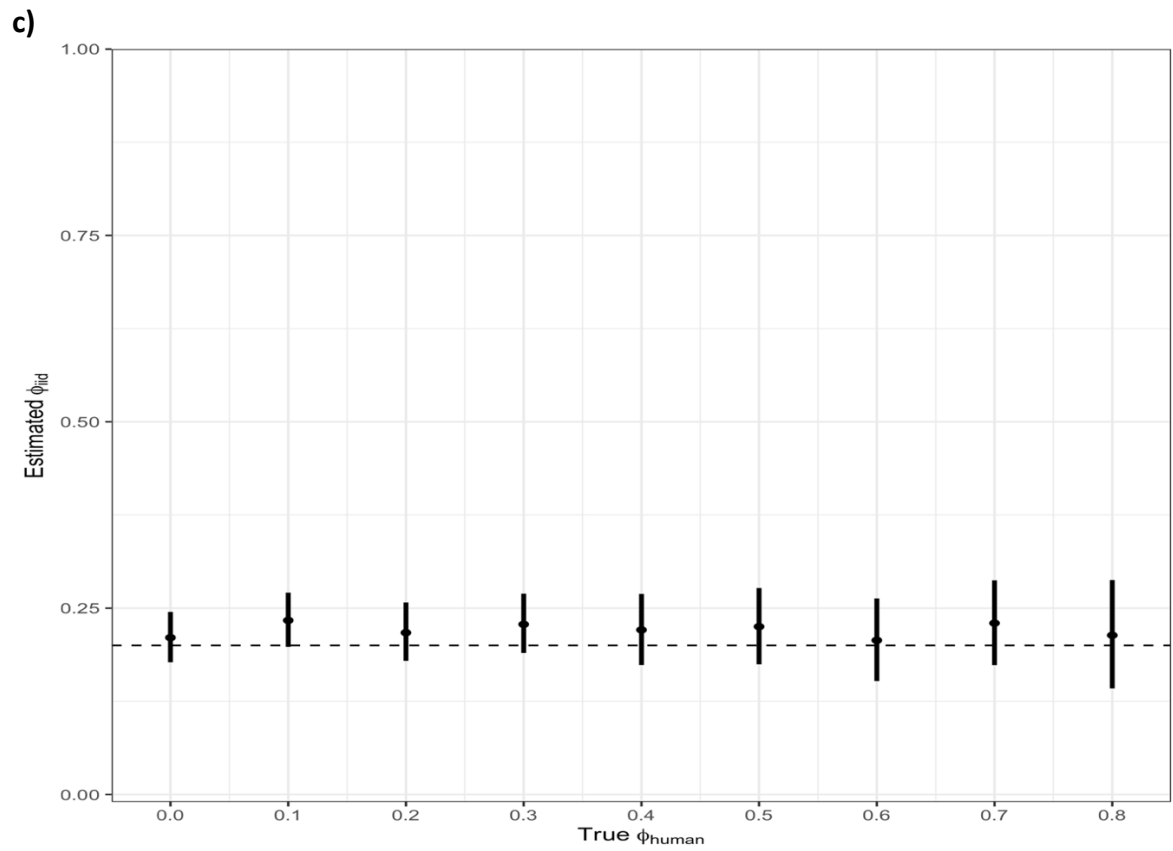

Figure M10: Mean and 95% credible interval of the proportion of variance of the random effects explained by a) the distance-based structure, b) the human movement-based structure, and c) unstructured random terms. Dashed lines represent the true value from simulations.

# References

1. Cox MA, Cox TF. 2008 Multidimensional scaling. In *Handbook of data visualization*, pp. 315–347. Springer.

2. Simini F, González MC, Maritan A, Barabási A-L. 2012 A universal model for mobility and migration patterns. *Nature* **484**, 96–100.

3. Tizzoni M, Bajardi P, Decuyper A, Kon Kam King G, Schneider CM, Blondel V, Smoreda Z, González MC, Colizza V. 2014 On the use of human mobility proxies for modeling epidemics. *PLoS Comput. Biol.* **10**, e1003716. (doi:10.1371/journal.pcbi.1003716)

4. Estatística IB de G e. 2020 *Regiões de influência das cidades 2018*. IBGE Rio de Janeiro.

5. Wood SN. 2006 Low-Rank Scale-Invariant Tensor Product Smooths for Generalized Additive Mixed Models. *Biometrics* **62**, 1025–1036. (doi:10.1111/j.1541-0420.2006.00574.x)

6. Wood S, Wood MS. 2015 Package 'mgcv'. *R Package Version* **1**, 729.

7. Wood SN. 2011 Fast stable restricted maximum likelihood and marginal likelihood estimation of semiparametric generalized linear models. *J. R. Stat. Soc. Ser. B Stat. Methodol.* **73**, 3–36. (doi:10.1111/j.1467-9868.2010.00749.x)

8. Wood SN. 2017 *Generalized additive models: an introduction with R*.

9. Wood SN. 2016 Just Another Gibbs Additive Modeler: Interfacing JAGS and mgcv. *J. Stat. Softw.* **75**, 1–15. (doi:10.18637/jss.v075.i07)

10. De Valpine P *et al.* 2021 Nimble: MCMC, particle filtering, and programmable hierarchical modeling. *R Package Version 011* **1**.

11. de Valpine P, Turek D, Paciorek CJ, Anderson-Bergman C, Lang DT, Bodik R. 2017 Programming with models: writing statistical algorithms for general model structures with NIMBLE. *J. Comput. Graph. Stat.* **26**, 403–413.

12. Rue H, Riebler A, Sørbye SH, Illian JB, Simpson DP, Lindgren FK. 2017 Bayesian computing with INLA: a review. *Annu. Rev. Stat. Its Appl.* **4**, 395–421.

13. Riebler A, Sørbye SH, Simpson D, Rue H. 2016 An intuitive Bayesian spatial model for disease mapping that accounts for scaling. *Stat. Methods Med. Res.* **25**, 1145–1165. (doi:10.1177/0962280216660421)

14. Simpson D, Rue H, Riebler AI, Martins TG, Sørbye SH. 2017 Penalising Model Component Complexity: A Principled, Practical Approach to Constructing Priors. (doi:10.1214/16-STS576)

15. Wood SN. 2017 *Generalized additive models: an introduction with R*. CRC press.

16. Martino S, Rue H. 2009 Implementing approximate Bayesian inference using Integrated Nested Laplace Approximation: A manual for the inla program. *Dep. Math. Sci. NTNU Nor.*

# Supplementary figures

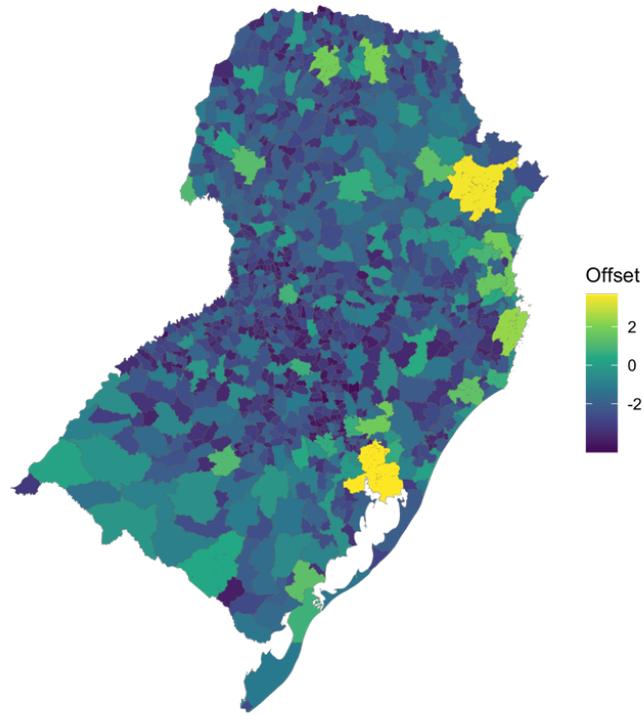

**Figure S1: Offset, $log(\xi_i)$, the log of the population of each municipality divided by 100,000 in South Brazil.**

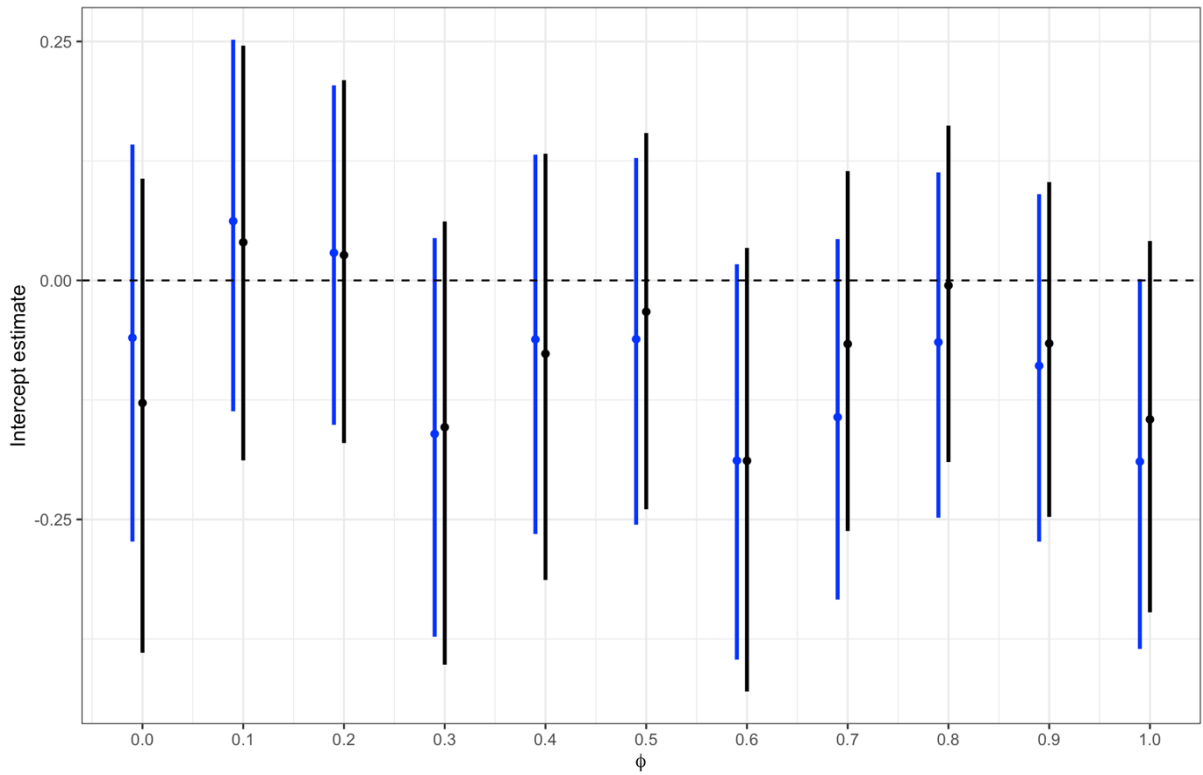

**Figure S2: Mean and 95% credible interval of the intercept coefficient estimates from the smoothing spline (black) and BYM2 (blue) models, compared to the true simulated value, 0.**

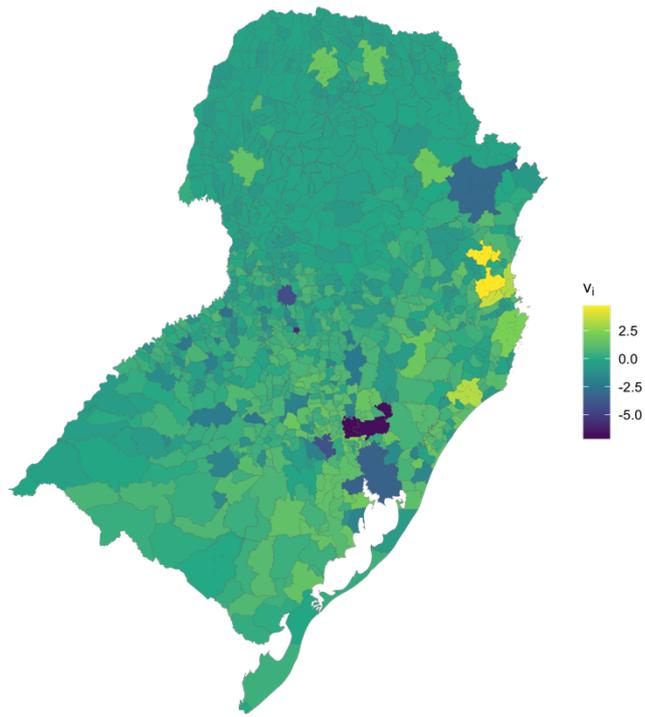

**Figure S3:** Human movement-based smooth function created by applying the function $sm(x_i, z_i)$ to coordinates describing the connectivity between municipalities in South Brazil arising from human movement.

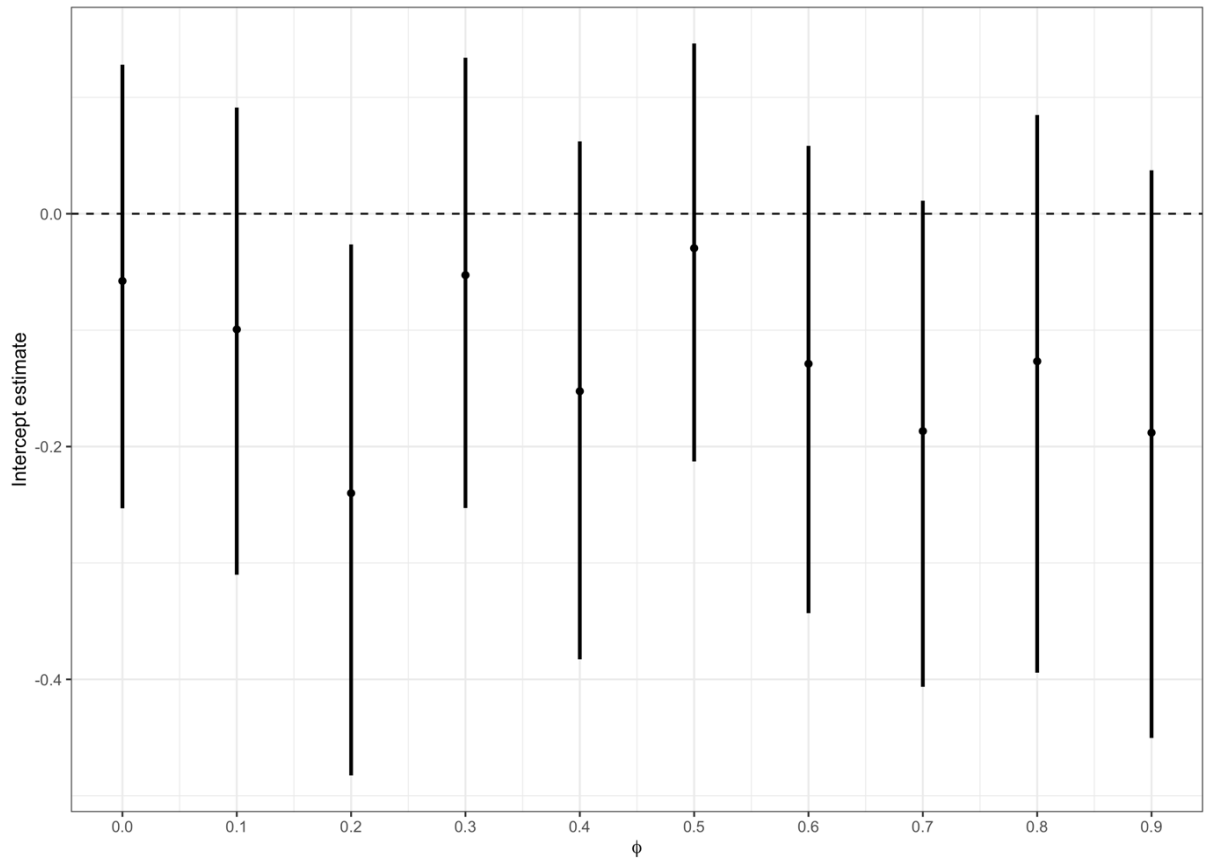

**Figure S4: Mean and 95% credible interval of the intercept coefficient estimates, compared to the true simulated value, 0 (sorted by $\phi_2$).**